\documentclass{article}
\usepackage{arxiv}
\usepackage[utf8]{inputenc} 
\usepackage[T1]{fontenc}    
\usepackage{url}            
\usepackage{booktabs}       
\usepackage{amsfonts}       
\usepackage{nicefrac}       
\usepackage{microtype}      
\usepackage{graphicx}
\usepackage[square,sort,comma,numbers]{natbib}
\usepackage{doi}
\usepackage{graphicx}
\usepackage{subfigure}
\usepackage{miller}
\usepackage{amsmath}
\usepackage{amssymb}
\usepackage{float}
\usepackage{color,soul}
\usepackage{mathtools}
\usepackage{appendix}
\usepackage{color}
\usepackage{tikz}
\usepackage{siunitx}
\usepackage{lineno}
\usepackage[normalem]{ulem}
\usepackage{upgreek}
\usepackage{siunitx}
\usepackage{fancyhdr}

\graphicspath{{figures/}}

\title{Orientation reconstruction of transformation \texorpdfstring{$\upalpha$}{alpha} titanium alloys via
polarized light microscopy: methodology and assessment}

\author{Amit Singh \\
	Department of Mechanical Engineering\\
	The University of Alabama\\
	Tuscaloosa, AL 35487 \\
	\texttt{asingh38@crimson.ua.edu.edu} \\
	\And
	Mark Obstalecki \\
	Air Force Research Laboratory\\
	Materials and Manufacturing Directorate\\
	Wright Patterson AFB, OH 45433 \\
	\texttt{mark.obstalecki@us.af.mil} \\
        \And
	Darren C. Pagan \\
	Department of Materials Science and Engineering\\
	The Pennsylvania State University\\
	University Park, PA 16802 \\
	\texttt{dcp5303@psu.edu} \\
        \And
	Michael Glavicic \\
	Rolls-Royce Corporation\\
	Indianapolis, IN 46225\\
	\texttt{michael.glavicic@rolls-royce.com} \\
        \And
	Matthew Kasemer \\
	Department of Mechanical Engineering\\
	The University of Alabama\\
	Tuscaloosa, AL 35487 \\
	\texttt{mkasemer@eng.ua.edu} \\
}

\renewcommand{\shorttitle}{Singh et. al. 2024}

\hypersetup{
pdftitle={Transformation \texorpdfstring{$\upalpha$}{alpha} titanium alloy orientation reconstruction via polarized light microscopy: methodology and assessment},
pdfsubject={q-bio.NC, q-bio.QM},
pdfauthor={Amit Singh, Mark Obstalecki, Darren C. Pagan, Michael Glavicic, Matthew Kasemer},
pdfkeywords={Polarized light microscopy, Titanium, Burgers orientation relationship, Crystal plasticity, Stress analysis},
}

\begin{document}
\maketitle

\begin{abstract}
Emerging microstructural characterization methods have received increased attention owing to their promise of relatively inexpensive and rapid measurement of polycrystalline surface morphology and crystallographic orientations. Among these nascent methods, polarized light microscopy (PLM) is attractive for characterizing alloys comprised of hexagonal crystals, but is hindered by its inability to measure complete crystal orientations. In this study, we explore the potential to reconstruct quasi-deterministic orientations for titanium microstructures characterized via PLM by considering the Burgers orientation relationship between the room temperature $\upalpha$ (HCP) phase fibers measured via PLM, and the $\upbeta$ (BCC) phase orientations of the parent grains present above the transus temperature. We describe this method---which is capable of narrowing down the orientations to one of four possibilities---and demonstrate its abilities on idealized computational samples in which the parent $\upbeta$ microstructure is fully, unambiguously known. We further utilize this method to inform the instantiation of samples for crystal plasticity simulations, and demonstrate the significant improvement in deformation field predictions when utilizing this reconstruction method compared to using results from traditional PLM.
\end{abstract}


\section{Introduction}
\label{sec:introduction}

Research into fundamental crystal-scale behavior has led to a decrease in uncertainty with respect to macroscopic mechanical behavior of polycrystalline materials, which has in turn facilitated the design of materials and engineering components which are able to meet increasingly stringent demands in modern machinery. Significant experimental developments were made in the latter half of the 20th century regarding a deeper understanding of how the morphology of grains and the collective distribution of crystallographic orientations (or: crystallographic texture) influence the behavior of materials, primarily in terms of the yield strength and ductility. In parallel, developments in microstructurally-sensitive deformation modeling techniques have similarly enjoyed significant progress. Various techniques have been developed in this regard, from mean field methods in the 1930s~\cite{sachs,taylor}, to self-consistent modeling primarily developed in the 1980s and 1990s~\cite{LEBENSOHN1993}. First developed in the late 1980s~\cite{asaro,nemat1986rate} as a method for modeling texture evolution, crystal plasticity finite element modeling (CPFEM) has evolved to a full-field method capable of capturing the intragrain deformation response of high-fidelity representations of microstructures (considering both crystallographic texture and explicit representations of complex grain morphology)~\cite{marin1,marin2,roters}, and has been proven adept at the prediction of both intragrain and macroscopic behavior, and has offered unparalleled insight into the relationship of elasticity and plasticity at those scales (and thus our best window into structure-property relationships).

In all of the above modeling methods, the models require some degree of experimental material characterization or micromechanical testing to aid in initiation, verification, or validation of the model. For CPFEM specifically, proper instantiation of a virtual polycrystalline sample that faithfully represents a physical sample requires (at least) information about the geometric morpohology of the grains, as well as the grains orientations (i.e., crystallographic texture)~\cite{roters}. Most often, two-dimensional characterization methods such as electron backscatter diffraction (EBSD) are used to measure both granular morphology and orientations on the surfaces of samples, and various methods have been developed to infer the three-dimensional morphology (either statistically~\cite{Takahashi2003,JAVAHERI2020}, based on oblique surface measurements~\cite{TURNER2016}, or destructively in the case of serial sectioning~\cite{echlinss}). More recently, high energy diffraction microscopy (HEDM) has offered the ability to non-destructively map orientations in three dimensions~\cite{suter2006forward,bernier2011far,pagan2018measuring,nygren2020algorithm,pagan2021analysis}, which can be used to directly construct digital twins of actual physical samples~\cite{Quey2018}, rather than statistically-representative virtual samples. Further, in the case of HEDM, micromechanical testing may be performed on the samples after initial characterization, allowing for unprecedented experimental deduction of structure-property relationships~\cite{Hurley2019,Chatterjee2019}.

Unfortunately, while EBSD and HEDM have become well-accepted as characterization techniques and are indeed powerful tools which have progressed significantly in terms of cost and ease of use, they suffer from practical limitations. Both are relatively time consuming (e.g., PLM can image large scans at a sub micron resolution in a fraction of the time---minutes versus hours---compared to EBSD or HEDM), expensive (e.g., EBSD is generally charged by the hour, and is thus naturally more expensive, while HEDM requires a synchrotron light source), and require a relatively high degree of training necessary to produce skilled researchers capable of performing proper, meaningful measurements compared to the method of PLM (this latter point is not meant to disparage these techniques, but rather highlight the inherent difficulty in the theory necessary to expertly employ them). Consequently, it is beneficial to develop methods which significantly reduce cost, lead-time, and/or the knowledge/skill barrier necessary to produce adequate characterization results, especially in cases where high-throughput is desired. Recently, significant research attention has been given to the development of ultrasonic~\cite{LAN2018,Dryburgh2020,LIU2021,He2022} and optical techniques in an effort to address these problems. Of interest to this study, we highlight the advent of polarized light microscopy (PLM) as a characterization method for materials comprised of crystals exhibiting hexagonal symmetry~\cite{Jin2020,hoover,Chao2024}. PLM offers the ability to measure spatial fields of (partial) crystal orientations both relatively rapidly and using relatively affordable equipment (compared to EBSD), as well as with fewer material preparation steps. However, PLM is limited in that it only measures an incomplete description of the orientation of HCP crystals, which limits its applicability as a characterization method for micromechanical simulations owing to the potential for high errors in predictions~\cite{vanWees2023}.

In this study, we outline a method to reconstruct full orientations of the crystals in $\upalpha$ titanium materials to one of four orientations---i.e., a quasi-deterministic approach. As prime inspiration for this study, we consult the work of Glavicic et. al.~\cite{Glavicic2003}, which established a method exploiting the physics of the allotropic phase transformation that many titanium alloys undergo during cooling in an effort to deduce the orientation field of the BCC ($\upbeta$ phase) microstructure that likely existed at high temperature based on the measurement of the room temperature $\upalpha$ phase orientation field. Here, we demonstrate that the measurement of $\upalpha$ colony \emph{crystallographic fibers} via PLM likewise allow for the back-calculation of the parent $\upbeta$ grain orientations to two possibilities, from which 24 deterministic $\upalpha$ colony orientations are forward-calculated. Of these available $\upalpha$ orientations, we show that four share a $c$ axis orientation necessary to satisfy that which PLM is able to measure. We demonstrate this reconstruction method on a synthetic sample where the parent $\upbeta$ phase microstructure is fully and unambiguously known (we opt for virtual samples to unambiguously validate our method rather than comparison to another method, which is prone to experimental uncertainty as well as sensitivity to reconstruction parameters). Similar to our previous study quantifying the consequences of PLM's orientation ambiguity in CPFEM predictions~\cite{vanWees2023}, we here demonstrate that the choice of one of the four possible orientations leads to significant improvement in deformation field predictions compared to what can be realized via traditional PLM.

\section{Background}
\label{subsec:background}

\subsection{\texorpdfstring{$\upalpha$/$\upbeta$}{alpha beta} Titanium Transformation}
\label{subsec:tibackground}

In pure titanium and many titanium alloys, the material undergoes an allotropic phase transformation during cooling, transforming from an exclusive body centered cubic structure (the $\upbeta$ phase), to primarily a hexagonal close packed structure (the $\upalpha$ phase), with the possibility of some fraction of retained $\upbeta$ phase (on the order of 10\% volume). In pure titanium, the temperature at which this transformation takes place, or the transus temperature, is approximately \SI{882}{\celsius}~\cite{Lutjering2003}. Generally, as the $\upbeta$ phase grains (herein: ``parent $\upbeta$ grains'') cool, they tend to transform into multiple distinct sub-grain regions of $\upalpha$ phase (herein: ``$\upalpha$ colonies''), which present as individual grains in the cooled microstructure, though collectively retain the approximate geometric morphology of their common parent $\upbeta$ grain (or, at least, the collection of $\upalpha$ colonies which descend from a common parent $\upbeta$ grain will be connected in a spatially contiguous network, if not perfectly adhering to the exact morphology of the parent $\upbeta$ grain). 

The orientations that these $\upalpha$ colonies may transform to are coupled to the orientation of their parent $\upbeta$ grain via the Burgers orientation relationship, or:
\begin{subequations}
    \label{eq:bor}
    \begin{align}
        \hkl{1 1 0}_{\upbeta}~&||~\hkl(0 0 0 1)_{\upalpha} \quad , \\
        \hkl<1 -1 1>_{\upbeta}~&||~\hkl<1 1 -2 0>_{\upalpha} \quad .
    \end{align}
\end{subequations}
Adherent to this relationship, we can calculate the possible $\upalpha$ colony orientations which may transform from a single parent $\upbeta$ grain orientation. Described in detail in~\cite{Glavicic2003}, the relationship takes the mathematical form:
\begin{equation}
    \label{eq:betatoalpha}
    {\bf S}_i^{\upalpha} {\bf R}^{\upalpha} = {\bf B} {\bf S}_j^{\upbeta} {\bf R}^{\upbeta} \quad , 
\end{equation}
where ${\bf S}_i^{\upalpha}$ refers to the 12 HCP symmetry operators and ${\bf S}_j^{\upbeta}$ refers to the 24 BCC symmetry operators, ${\bf B}$ to a rotation derived from the Burgers orientation relationship, and ${\bf R}^{\upalpha}$ and ${\bf R}^{\upbeta}$ are the $\upalpha$ colony and parent $\upbeta$ grain orientations, respectively, parameterized as rotation matrices. For a single $\upbeta$ grain orientation, this relationship yields 12 unique $\upalpha$ colony orientations that may arise due to the transformation, also known as $\upalpha$ variants. Conversely, we may calculate the possible parent $\upbeta$ grain orientations that a given $\upalpha$ colony orientation may have transformed from, yielding six unique parent $\upbeta$ grain orientations.

In~\cite{Glavicic2003}, a method is described in which the coupling between the $\upalpha$ and $\upbeta$ orientations are used to determine the orientation field of the parent $\upbeta$ grain structure that likely existed above the transus temperature given a field of measured $\upalpha$ orientations. This is achieved via the consideration of spatial location/contiguity of measured $\upalpha$ colonies who have collective intersect among their possible parent $\upbeta$ variants. Ideally, a collection of contiguous $\upalpha$ colonies intersect at only one parent $\upbeta$ variant, allowing for unambiguous determination of the parent $\upbeta$ grain orientation. This method has been demonstrated on two-dimensional orientation fields (e.g., via  electron backscatter diffraction (EBSD)~\cite{Glavicic2003}), as well as three-dimension orientation fields (e.g., via HEDM~\cite{Wielewski2015,Kasemer2017}), leading to plausible results regarding the possible $\upbeta$ orientation field that existed above the transus temperature.

\subsection{Measurement of \texorpdfstring{$\upalpha$}{alpha} Fibers via PLM}
\label{subsec:fibermeasure}

Polarized light microscopy (PLM) is a well-established optical method used to image the microstructure of metallic materials by exploiting their anisotropic refractive properties. In essence, the crystallographic orientations of grains relative to a light source determines the intensity of refraction, allowing for imaging of microstructural features. Overall, PLM is considered to be relatively low-cost and rapid compared to conventional two-dimensional characterization methods such as EBSD. Regarding the relative speed, this facilitates the imaging of much larger areas of materials that would otherwise be inaccessible or intractable via EBSD.

More recently, it has been shown that the measurement of the field of refracted intensities while rotating a sample comprised of HCP crystals allows for the determination of the field of $c$ axis orientations (described in detail in~\cite{Jin2020}). While promising as a method for orientation determination, the prime limitation of PLM in this regard is its inability to determine the full orientation of the crystal. Specifically, PLM is unable to discern:
\begin{enumerate}
    \item between any of the orientations with the same $c$ axis orientation, or
    \item the true orientation of the $c$ axis due to periodicity in measured intensity.
\end{enumerate}
We discuss each of these ambiguities separately below.

\subsubsection{Ambiguity of Rotation About the \texorpdfstring{$c$}{c} Axis}
\label{subsubsec:caxisambiguity}

Discussed in detail in~\cite{Jin2020}, the anisotropic nature of optical refraction in HCP materials renders PLM unable to discern the amount of rotation about the $c$ axis. This may be best understood via consideration of the Euler-Bunge orientation parameterization~\cite{bunge}: $\left(\phi_1, \Phi, \phi_2\right)$, or rotations about the $z$, $x^\prime$, $z^{\prime\prime}$ axes, respectively. In the Euler-Bunge parameterization, $\phi_1$ and $\Phi$ determine the orientation of the $c$ axis, while $\phi_2$ determines the rotation about the $c$ axis. PLM, thus, is able determine only $\phi_1$ and $\Phi$, while $\phi_2$ is left ambiguous.

Alternatively, the abilities of PLM may be understood as the measurement of \emph{crystallographic fibers}, or specifically the sets of orientations which share the same $c$ axis rather than deterministic orientations. Such a fiber may be defined as the set of orientations, ${\bf R}$ (parameterized as a rotation matrix in a convention transforming vectors in the crystal basis to the sample basis), which satisfy:
\begin{equation}
    \label{eq:cfiber}
    {\bf R} \begin{Bmatrix}
        0 \\
        0 \\
        1
    \end{Bmatrix} = \pm {\bf s} \quad ,
\end{equation}
where ${\bf s}$ is a normalized Cartesian vector representing the direction of the $c$ axis in the sample frame. Similar to the above ambiguity in the Euler-Bunge angles, here the rotation matrix does not need to be fully defined: only the third column of the matrix must be known to satisfy the above condition (and is equal to ${\bf s}$).

\begin{figure}[htbp!]
    \centering
    \subfigure[]{%
	\includegraphics[height=0.325\textwidth]{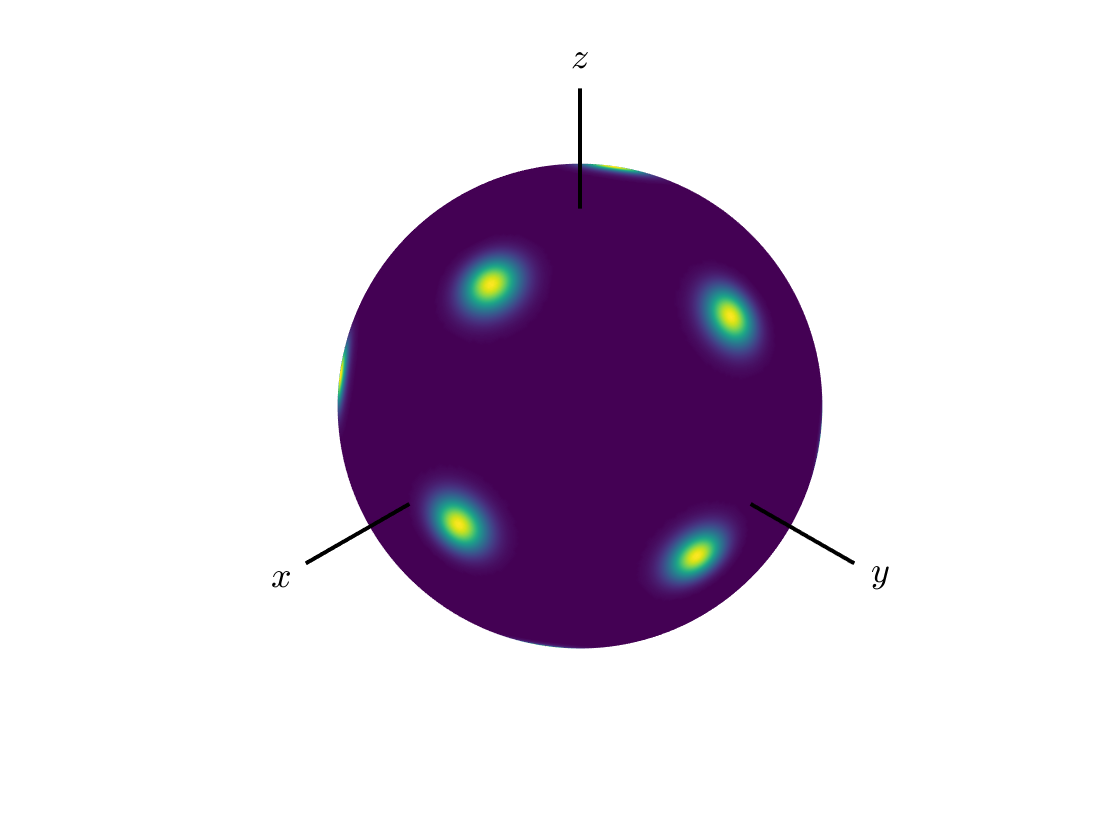}
        \label{subfig:alphapf}}
    \subfigure[]{%
	\includegraphics[height=0.325\textwidth]{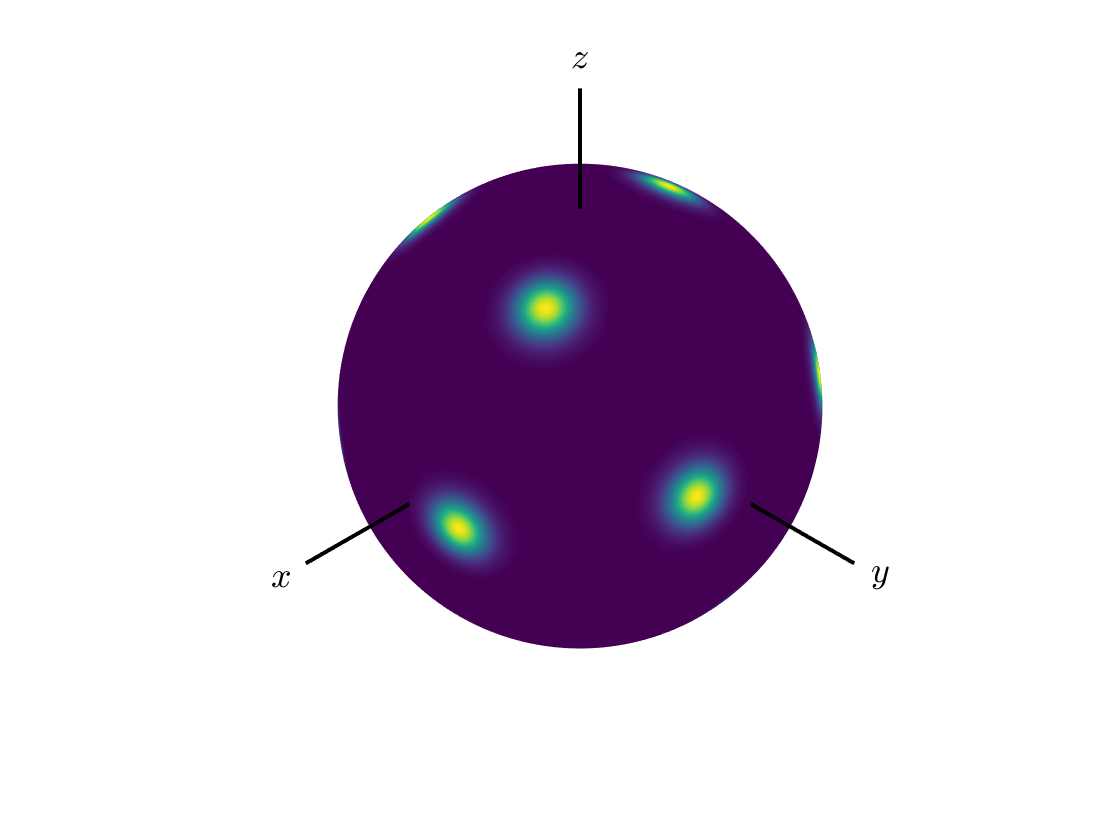}
        \label{subfig:alphapf180}}
    \caption{\subref{subfig:alphapf} A $\hkl{0 0 0 1}_\upalpha$ pole figure depicting the $c$ axes of the $\upalpha$ colonies that may descend from a single randomly-oriented parent $\upbeta$ grain, and \subref{subfig:alphapf180} a $\hkl{0 0 0 1}_\upalpha$ pole figure depicting the reflected $c$ axes that arise from the ambiguity due to the periodicity in measured intensities.}
    \label{fig:alphapfs}
\end{figure}

We can compactly represent the fibers that PLM is able to measure on pole figures. As demonstration, we present a $\hkl{0 0 0 1}_\upalpha$ pole figure in Figure~\ref{subfig:alphapf}, depicting the orientations of the $c$ axes for the $\upalpha$ colonies that may descend from a single parent $\upbeta$ grain with a random orientation (as calculated following the description in Section~\ref{subsec:tibackground}). We note the presence of 6 unique peaks (along with 6 symmetrically equivalent negative peaks that appear on the reverse of the pole figure, not shown). This overall indicates that of the 12 possible $\upalpha$ colony orientations from a single parent $\upbeta$ grain orientation, two each share the same $c$ axis orientation---i.e., two $\upalpha$ colony orientations lie along the same crystallographic fiber (we provide a demonstration of this property via consideration of the Rodrigues parameterization in Appendix~\ref{subsec:demonstration}).

\subsubsection{Ambiguity of \texorpdfstring{$c$}{c} Axis Orientation Due to Periodicity in Measured Intensity}
\label{subsubsec:caxisambiguity2}

For PLM characterization of HCP specimens, the light intensity data that is collected as the specimen is rotated in relation to the polarization filter displays a periodicity of \SI{180}{\degree} (as demonstrated in~\cite{Jin2020}). As a result of this \SI{180}{\degree} periodicity in measured intensities, there exists two $c$ axis orientations which cannot be differentiated via PLM~\footnote{We note that similar ambiguity exists in ultrasonic characterization methods, particularly spatially resolved acoustic spectroscopy (SRAS), where the observed periodicity is due to the measured directional wave velocity's proportionality to the anisotropic elastic moduli~\cite{Dryburgh2020}.}. Inspecting the consequences on the Euler-Bunge angles, assuming that the sample in question is rotated about the sample $z$ direction during data collection, the \SI{180}{\degree} periodicity affects the measurement of $\phi_1$---the measured angle will either be correct, or potentially a mis-indexing of the true orientation of $\phi_1\pm\SI{180}{\degree}$.

To demonstrate the consequences of the second ambiguity on the fibers of $\upalpha$ colony $c$ axis orientations presented in Figure~\ref{subfig:alphapf}, we calculate a new set of $\upalpha$ colony orientations by altering $\phi_1$ in the above-described manner. Herein, we refer to these as the ``reflected'' orientations, which we present in Figure~\ref{subfig:alphapf180} for this particular example. We contend that PLM is unable to determine \emph{either of the pole figures} in Figure~\ref{fig:alphapfs} as containing the true set of $c$ axes versus their reflections, due to the periodic nature of measured intensities (again, we provide an alternative depiction of the consequences of this ambiguity on the $\upalpha$ colony fibers and orientations plotted in Rodrigues space in \ref{subsec:demonstration}).

\section{Methodology for Reconstruction of Quasi-Determinstic \texorpdfstring{$\upalpha$}{alpha} Orientations}
\label{sec:methodology}

Having established that PLM is able to measure $\upalpha$ colony fibers as opposed to unambiguous orientations, we now present a methodology to reconstruct the $\upalpha$ colony orientations to one of four possibilities---i.e., a quasi-deterministic approach. Our methodology is inspired by the approach utilized in~\cite{Glavicic2003} in which the parent $\upbeta$ grain orientations likely present above the transus temperature may be calculated utilizing the deterministic $\upalpha$ colony orientations. Overall, the flow of our method is summarized as follows:
\begin{enumerate}
    \item Cluster $\upalpha$ colonies as having a shared parent $\upbeta$ grain lineage based on the relative misorientation between their $c$ axes and their spatial location.
    \item Process a $\hkl{0 0 0 1}_\upalpha$ pole figure for each $\upbeta$ grain representing all measured $\upalpha$ colony $c$ axes within the parent $\upbeta$ grain and their reflections.
    \item Back-calculate the possible parent $\upbeta$ grain orientations by registering the processed $\hkl{0 0 0 1}_\upalpha$ pole figure against entries in a dictionary containing $\hkl{0 0 0 1}_\upalpha$ pole figure of known parent $\upbeta$ grain lineage. Two parent distinct $\upbeta$ grain orientations will register.
    \item Forward-calculate the 24 deterministic $\upalpha$ colony orientations (12 each from each parent $\upbeta$ grain orientation).
    \item For each $\upalpha$ colony within a parent $\upbeta$ grain, find the four $\upalpha$ colony orientations that share a $c$ axis or reflected $c$ axis with the measured value.
\end{enumerate}
In this section, we present a detailed discussion of the above-summarized method. We further present an alternative demonstration in Rodrigues space in Appendix~\ref{sec:methodologyrod}.

\subsection{Processing of Measured Data}

We begin by processing the measured data in an effort to generate $\hkl{0 0 0 1}_\upalpha$ pole figures for each set of $\upalpha$ colonies that we determine to share the same parent $\upbeta$ grain lineage. The challenges here are two-fold: namely that we must determine what $\upalpha$ colonies share the same parent $\upbeta$ grain lineage, but that we are also unsure that the $c$ axes we measure are true or reflected.

\subsubsection{Calculation of Reflected \texorpdfstring{$c$}{c} Axes}
\label{subsubsec:reflected}

We continue by processing the measured $c$ axes for later use in clustering to determine shared parent $\upbeta$ lineage. We note that we are uncertain whether the $c$ axes we measure belong to the set of true peaks or to the reflected peaks, owing to the ambiguity discussed in Section~\ref{subsubsec:caxisambiguity2}. It is conceivable---and even likely---that the measured $c$ axis orientation field will contain $c$ axes which belong to both sets. Consequently, we generate a second $c$ axis orientation for each of the measured $\upalpha$ colonies by rotating \SI{180}{\degree} about the surface normal (or, operationally, adding \SI{180}{\degree} to the measured $\phi_1$ value, assuming that the sample is rotated about its $z$ axis during measurement). We are left with two possible $c$ axis orientations for each $\upalpha$ colony, one representing the measured orientation, the other the reflection (again, which one is true remains ambiguous).

\subsubsection{Clustering \texorpdfstring{$\upalpha$}{alpha} Colonies Based on Shared \texorpdfstring{$\upbeta$}{beta} Lineage}
\label{subsubsec:clusteralgo}

To calculate whether a set of $\upalpha$ colonies share a common parent $\upbeta$ grain lineage, we must know both their relative spatial location, as well as whether they share a common parent $\upbeta$ grain orientation. 

Regarding the former, we assume that $\upalpha$ colonies that share a common parent $\upbeta$ grain lineage are spatially contiguous. To efficiently test for spatial contiguity of $\upalpha$ colonies, we construct a graph representation of a polycrystal, where the colonies in a polycrystal are represented as ``nodes'' in the graph, and the graph ``edges'' describe the connection between nodes, ultimately providing a simplified spatial map of the polycrystal~\cite{pagan}. Identification of colonies and their neighbors to facilitate construction of the graph follows standard grain indexing practices, such as with EBSD data. We represent the graph of the polycrystal as a (generally sparse) adjacency matrix, whose row and column indices refer to the indexed IDs of the $\upalpha$ colonies, and entries of ``1'' indicate connectivity or sharing a graph edge (i.e., a neighboring / spatially contiguous colony), and entries of ``0'' indicate a non-neighboring colony, allowing for efficient identification of neighborhood.

Regarding the latter, we do not know \emph{a priori} what the parent $\upbeta$ grain orientation is from which each $\upalpha$ colony descended. However, we can show that any two $\upalpha$ colony variants from a common parent $\upbeta$ grain lineage have $c$ axes that are either \SI{0}{\degree}, \SI{60}{\degree}, or \SI{90}{\degree} from one another. Thus, we have at our disposal an efficient method to down-select whether two contiguous $\upalpha$ colonies may have descended from a common parent $\upbeta$ grain lineage. For each $\upalpha$ colony, we compare either of its $c$ axes (measured or reflected) against both of a neighboring colony's $c$ axes (measured and reflected). If either of these values matches the above misorientation criteria, we can select the two colonies as sharing a common parent $\upbeta$ grain lineage. We recognize that both the transformation from $\upbeta$ to $\upalpha$ may be imperfect and PLM measurements contain uncertainty, and as such we cannot down-select $\upalpha$ colonies with rigid consideration of the angle between their $c$ axes. Instead, we allow for a $\pm$\SI{5}{\degree} tolerance.

Algorithmically, we start at any point of the graph and work neighbor-to-neighbor through the graph in a routine ``flood fill'' scheme~\cite{Wielewski2015}. In essence, we work through the graph testing (naturally) spatial contiguity of the $\upalpha$ colonies, as well as whether each neighboring $\upalpha$ colony shares a common parent $\upbeta$ grain lineage based on the above-described criteria comparing the angle between the $c$ axes. If two $\upalpha$ colonies satisfy both of these criteria, we record that they must share the same parent $\upbeta$ grain ID. After completely filling through the graph, we are left with a field of parent $\upbeta$ grain IDs assigned to each $\upalpha$ colony, or (in other words) networks of $\upalpha$ colonies belonging to the same parent $\upbeta$ grain. Further, for each parent $\upbeta$ grain, we generate a $\hkl{0 0 0 1}_\upalpha$ pole figure considering all of the measured and reflected $c$ axes for every $\upalpha$ colony determined to share the same parent $\upbeta$ grain lineage. We normalize the pole figure such that the integrated surface intensity is 1. We use this pole figure to register against a dictionary of pole figures to determine the parent $\upbeta$ grain orientations. 

\subsection{Registration of Measured Pole Figure Against Dictionary Entries}
\label{subsubsec:registration}

To find the parent $\upbeta$ grain orientations from which the $\upalpha$ colonies may have descended, we compare the measured/processed pole figure representing the measured and reflected $c$ axes of the colonies from a single parent $\upbeta$ grain against a dictionary of $\hkl{0 0 0 1}_\upalpha$ pole figures for colonies from known parent $\upbeta$ grain orientations.

\begin{figure}[htbp!]
    \centering
    \includegraphics[height=0.325\textwidth]{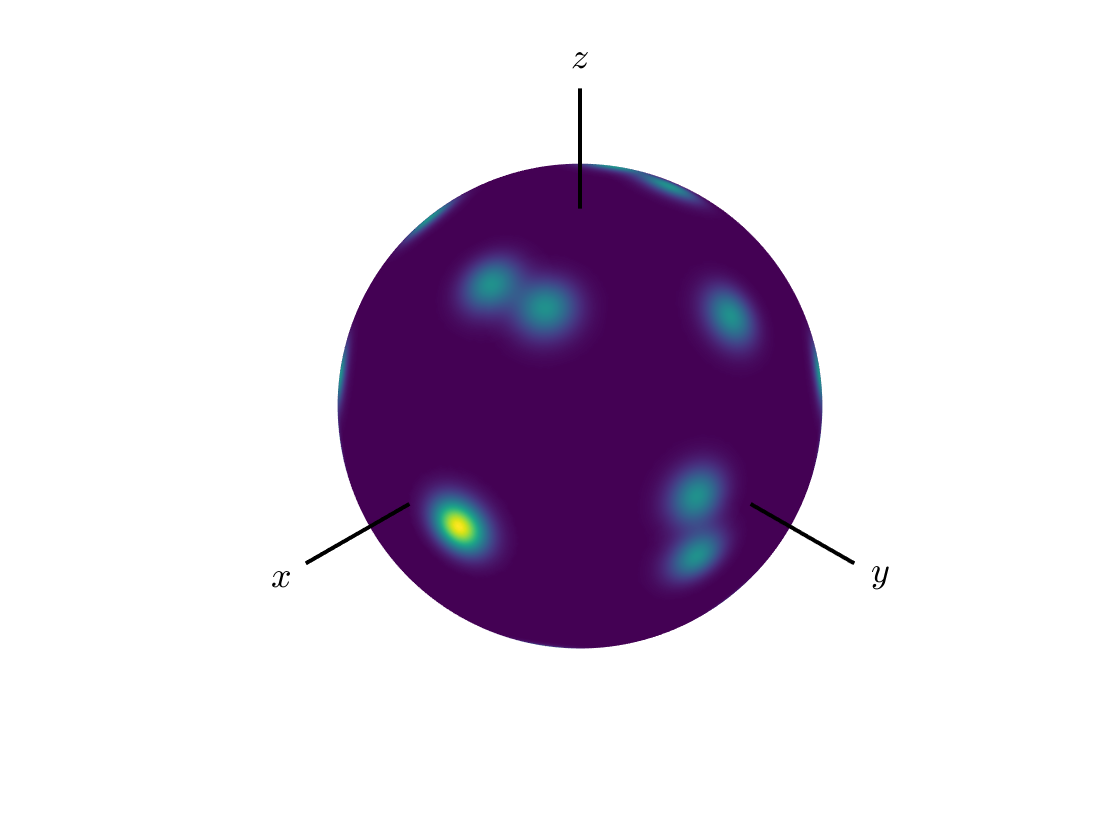}
    \caption{A $\hkl{0 0 0 1}_\upalpha$ pole figure depicting a typical dictionary entry, constructed of both the $c$ axes of the $\upalpha$ colonies that may descend from a single randomly-oriented parent $\upbeta$ grain, as well as their reflections.}
    \label{fig:alphapfmerged}
\end{figure}

We next generate a set of idealized $\hkl{0 0 0 1}_\upalpha$ pole figures representing the possible true and reflected $c$ axes that descend from parent $\upbeta$ grains of known orientation, termed a `dictionary'. To achieve this, we start by generating a list of parent $\upbeta$ grain orientations. We utilize the nodes of a finite element mesh of the cubic symmetry fundamental region of Rodrigues space~\cite{kumar} to generate a list of unique orientations (i.e., no symmetrically equivalent orientations) with a dense packing across the fundamental region (i.e., providing fine resolution of the parent $\upbeta$ grain orientations). For each of these parent $\upbeta$ grain orientations, we calculate the orientation of the $c$ axes of their $\upalpha$ colony variants as well as their reflections. We then construct a $\hkl{0 0 0 1}_\upalpha$ pole figure based on the set of unique $c$ axes, assuming a surface Gaussian distribution of minimal spread (average \SI{5}{\degree}) at each point to later aid in registration (this allows for some mismatch in registration in case of potential misaligned $c$ axes, due to imperfect transformation or uncertainty in measurement). Again, we normalize the pole figures such that the integrated surface intensity is 1. Figure~\ref{fig:alphapfmerged} depicts a typical dictionary entry (itself a summation of the true and reflected pole figures depicted previously in Figure~\ref{fig:alphapfs}).

With a dictionary of idealized pole figures, we register a measured pole figure against the dictionary entries to find the best match. We complete this by calculating the root mean squared error between the each dictionary entry and the measured pole figure. The dictionary entry with the smallest error is assumed to have the distribution of $c$ axes most similar to the measured set. We collect the orientation of the parent $\upbeta$ grain orientation corresponding to the idealized dictionary entry, and assign this orientation as the likely parent $\upbeta$ grain orientation for the set of measured $\upalpha$ colonies. We note that two separate $\upbeta$ orientations ultimately lead to the same pole figure dictionary entry, and thus we can determine the parent $\upbeta$ grain orientations to one of two values (demonstrated rigorously in Appendix~\ref{subsec:calcbetaori}).

\subsection{Forward Calculation of Quasi-Determinstic \texorpdfstring{$\upalpha$}{alpha} Colony Orientations}
\label{subsubsec:forwardalpha}

Knowing the two parent $\upbeta$ grain orientations which result in the same orientations of $\upalpha$ colony $c$ axes and their reflections, we calculate the true $\upalpha$ colony orientations based on the Burgers orientation relationship. For each of the two parent $\upbeta$ grain orientations, we calculate 12 unique $\upalpha$ colony orientations (24 total). We can show that among the 12 orientations from each $\upbeta$ grain, two each share the same $c$ axis orientation and cannot be further distinguished via PLM (see: ~\ref{subsec:demonstration}). Further, there exist two reflected orientations that share the reflected $c$ axis orientation of the measured, which again we cannot distinguish between via PLM. We are thus able to narrow down the possible $\upalpha$ colony orientations to one of four values.

\subsection{Experimental Complications}
\label{subsec:limitations}

We focus this study on the theoretical establishment of the technique presented above, and (to be discussed in Section~\ref{sec:cpsims}) the consequences when utilized in crystal plasticity finite element simulations, rather than on the experimental method of polarized light microscopy. There exists, however, some practical hindrances in terms of the experimental employ of this method that must be addressed. First, we note that an experimental sample must have proper surface preparation---similar to EBSD---such that the signal to noise ratio is maximized, providing confidence in the measurement of the field of $c$ axes. Failure to properly prepare the surface of the sample will increase uncertainty in (partial) orientation indexing, thus increasing uncertainty in the method presented in this study. Second, we note that we have thus far assumed that each $\upalpha$ colony variant for a given parent $\upbeta$ grain is present, such that the determination of the parent $\upbeta$ grains is limited to two orientations. If only one $\upalpha$ colony orientation from a given parent $\upbeta$ grain is present, then the parent $\upbeta$ grain orientation can only be determined to a fiber. This quickly collapses to only two orientations---particularly when considering the necessary reflections of the $\upalpha$ colonies. While we do not envision that there will exist a large fraction of parent $\upbeta$ grains which transform wholly to a single $\upalpha$ colony, there is a larger probability of edge cases---particularly along the physical boundary of the measured sample---where only a single $\upalpha$ colony variant from a parent $\upbeta$ grain is present, and thus confidence along the boundary of the sample measurement will be lower. Finally, we acknowledge that the Burgers orientation relationship may lead to imperfect transformations (i.e., ${\bf B}$ of Equation~\ref{eq:betatoalpha} may vary slightly), which may further introduce uncertainty into the reconstruction.

Collectively, an experimental implementation would likely be accompanied with a confidence metric to quantify the certainty of the orientation reconstruction. Motivated primarily by the facts that measurements may contain noise, and that transformations may be imperfect, we note that the pole figures constructed from measured poles and their reflections (similar to that presented in Figure~\ref{fig:alphapfmerged}) may contain peaks that are not perfect ``points'' on the pole figure, but rather ``spreads''. Since the dictionary entries are likewise constructed from peaks with a nominal degree of spread, this allows for imperfect mismatch between what is measured and what exists in the dictionary. We envision that a confidence metric could be constructed locally (i.e., as a field), and quantify (for example) the degree of mismatch between the measured peaks at that location and those in the chosen dictionary entry, likely via a misorientation metric. Other similar metrics could be borrowed from EBSD and appropriately modified for the case here. We do not pursue the construction of such a metric in this study, as we focus instead on the establishment and verification of the method via the idealized theoretical problem (as will be further motivated in Section~\ref{sec:demonstration}).

\section{Polycrystal Demonstration}
\label{sec:demonstration}

To demonstrate, we will utilize a virtual polycrystalline sample in which we generate a $\upbeta$ to $\upalpha$ transformation microstructure. While we envision that this method can and will be applied to experimental data, we opt to focus this study on the establishment of the method via the inspection of the idealized problem---i.e., one in which the sample contains $\upalpha$ colonies of only a single orientation, where the transformation is perfect, and there exists multiple $\upalpha$ colony variants within each parent $\upbeta$ grain. We note that while similar parent $\upbeta$ grain orientation fields may be reconstructed from EBSD data using commercial EBSD software, these fields are inferred rather than directly experimentally observed, and thus comparison against these results---which are like any other experimental data prone to uncertainty and sensitivity to choice of reconstruction parameters---is subpar when there exists an alternative that is perfect and unambiguous. The construction of virtual samples enables us to deterministically validate our method and software against known (true) deterministic orientations in which the field of both the room temperature $\upalpha$ colonies and the parent $\upbeta$ grains are both fully and unambiguously known.

\begin{figure}[htbp!]
    \centering
    \subfigure[]{%
	\includegraphics[height=0.325\textwidth]{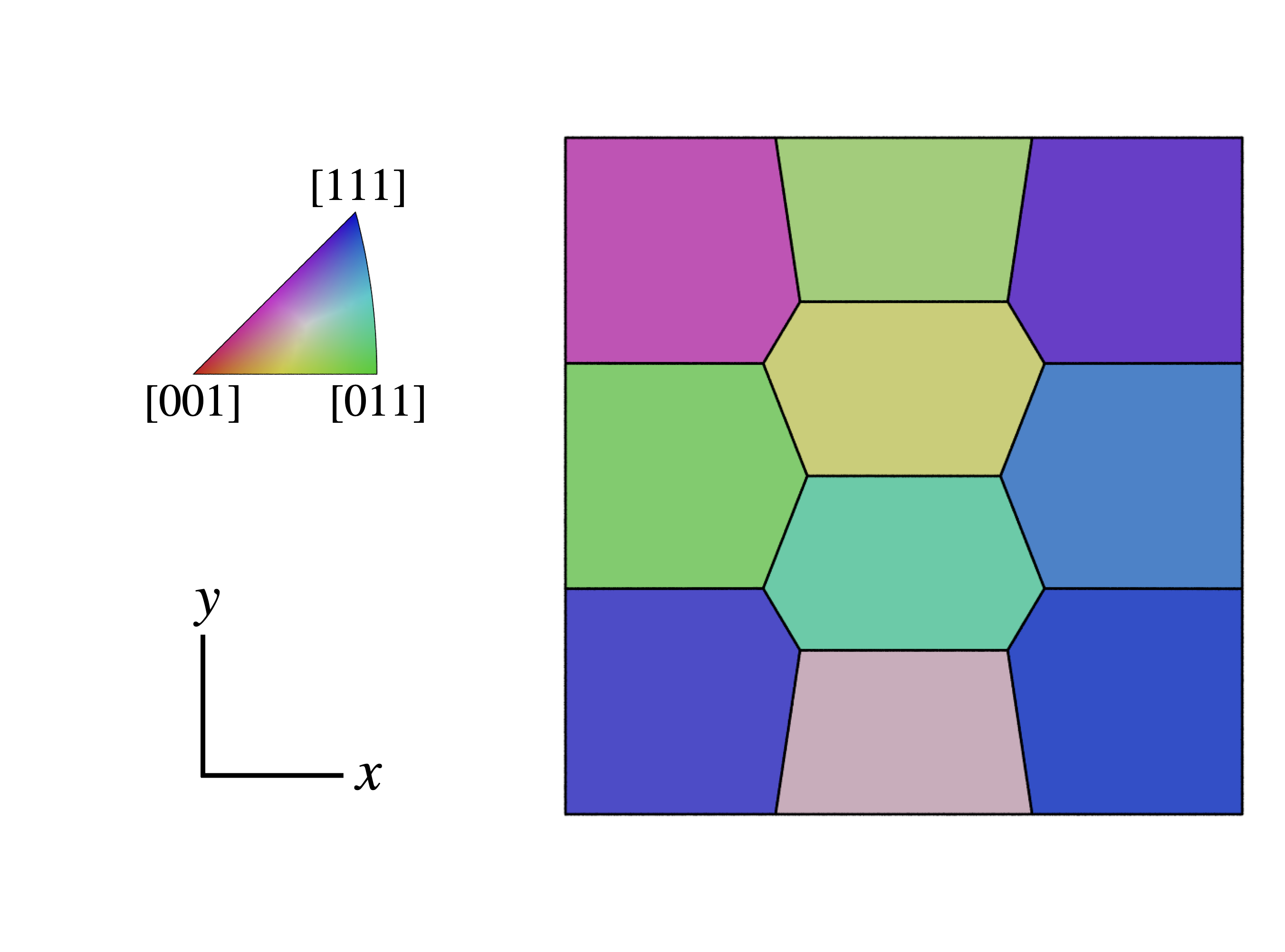}
        \label{subfig:2dbeta}}
    \subfigure[]{%
	\includegraphics[height=0.325\textwidth]{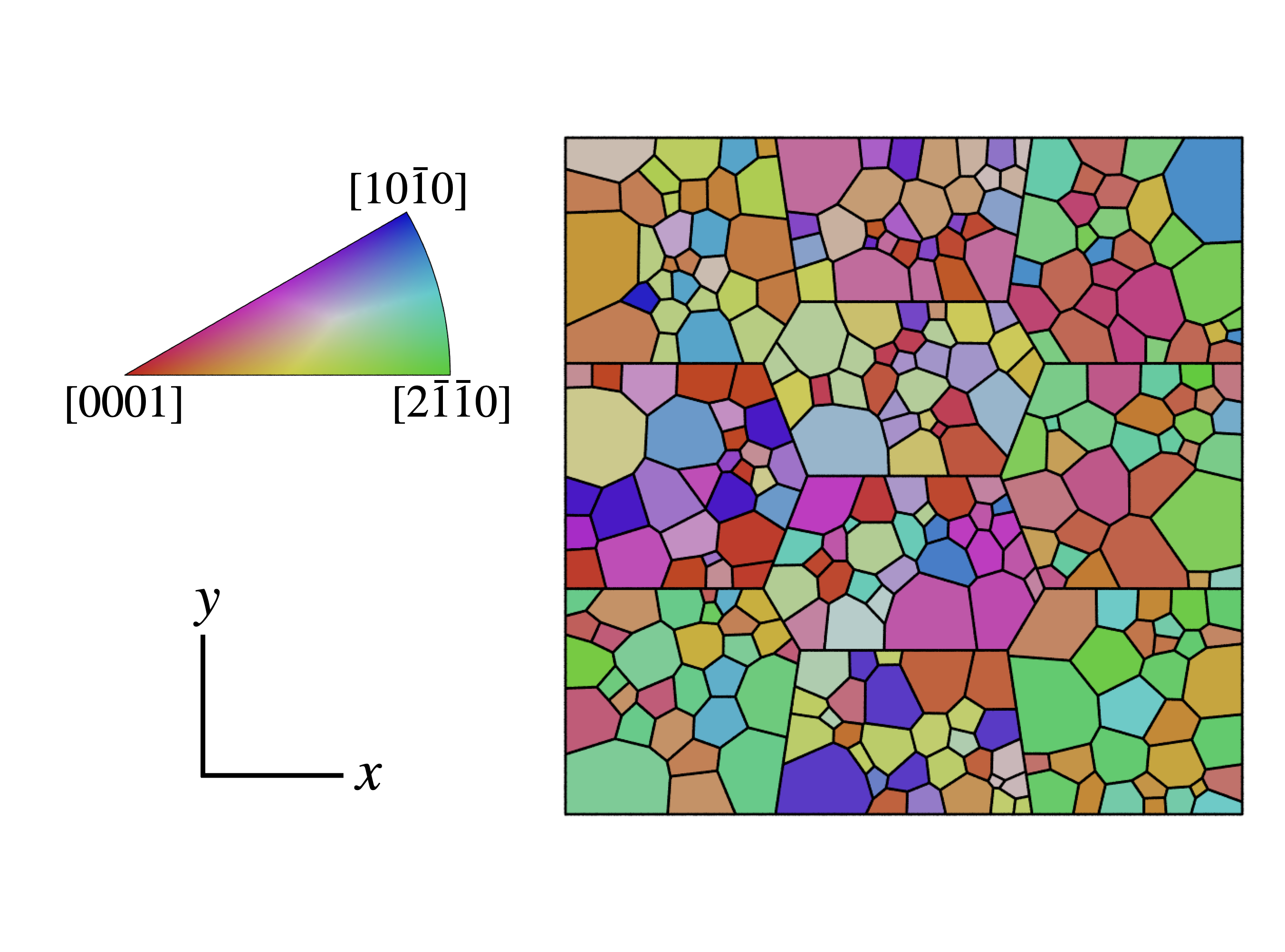}
        \label{subfig:2dalpha}}
    \caption{An idealized two-dimensional polycrystalline sample, representing \subref{subfig:2dbeta} the parent $\upbeta$ grain orientations, and \subref{subfig:2dalpha} the resulting $\upalpha$ colony orientations, which collectively serve as the starting point for the demonstration of the quasi-deterministic reconstruction method as employed on a polycrystalline domain. The inverse pole figure map is plotted with respect to the sample $z$ axis (i.e., out of page).}
    \label{fig:2dbetaalpha}
\end{figure}

To begin, we construct an idealized two-dimensional tessellation representative of a field of $\upalpha$ colonies with orientations adherent to a transformation structure (note: the tessellation method we utilize here is relatively unimportant for this demonstration, but is described in more detail later in Section~\ref{subsubsec:virtualsample}). We present a depiction of both the parent $\upbeta$ grains and the resulting $\upalpha$ colonies in Figure~\ref{fig:2dbetaalpha}. In this way, we begin with a sample of fully known $\upalpha$ colony orientations, as well as unambiguous knowledge of the parent $\upbeta$ grain microstructure. For sake of demonstration, we begin by altering the orientations to mimic what would be expected to be measured for such a microstructure via PLM. We achieve this by zeroing-out the third Euler-Bunge angle, $\phi_2$, for each $\upalpha$ colony orientations (thus enacting the first orientation ambiguity, Section~\ref{subsubsec:caxisambiguity}), as well adding (randomly, colony-to-colony) \SI{180}{\degree} to the first Euler-Bunge angle, $\phi_1$ (enacting the second orientation ambiguity, Section~\ref{subsubsec:caxisambiguity2}), ultimately rendering a set of orientations representative of those which could presumably be measured via PLM for the true virtual microstructure.

\begin{figure}[htbp!]
    \centering
    \subfigure[]{%
	\includegraphics[height=0.325\textwidth]{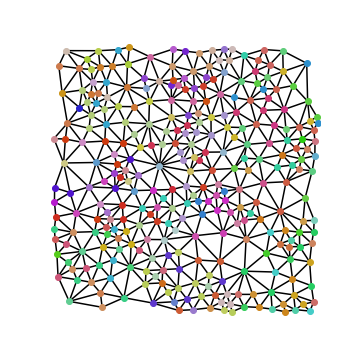}
        \label{subfig:alphagraph}}
    \subfigure[]{%
	\includegraphics[height=0.325\textwidth]{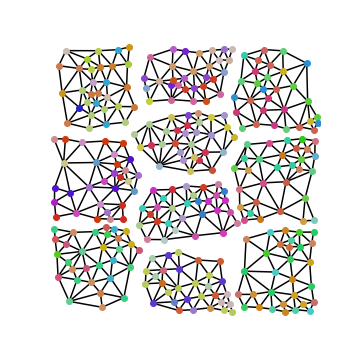}
        \label{subfig:betagraph}}
    \caption{\subref{subfig:alphagraph} A graph representation of the field of $\upalpha$ colonies shown in Figure~\ref{subfig:2dalpha}, where the nodes of the graph (dots) are plotted at the spatial location of the $\upalpha$ colony centroids, while the edges of the graph (lines) depict neighbor-to-neighbor connections, and~\subref{subfig:betagraph} the results of the flood fill algorithm, depicting the networks of $\upalpha$ colonies clustered by likely shared parent $\upbeta$ grain lineage (compare to Figure~\ref{subfig:2dbeta}).}
    \label{fig:graph}
\end{figure}

Next, we construct a graph representation of the field of $\upalpha$ colonies, which we present in Figure~\ref{subfig:alphagraph}. Each node of the graph represents a measured $\upalpha$ colony, while each line on the graph represents neighbor-to-neighbor connections. We note that at each node, we assign nodal attributes (i.e., descriptors of the $\upalpha$ colonies). For the sake of clustering the $\upalpha$ colonies based on their shared parent $\upbeta$ grain lineage, we assign both the vector representing the $c$ axis in the spatial frame, ${\bf s}$ (Equation~\ref{eq:cfiber}), as well as its reflected $c$ axis (as described in Section~\ref{subsubsec:reflected}). We then employ the flood-fill algorithm (Section~\ref{subsubsec:clusteralgo}) comparing the $c$ axes between each node and its neighbors (working node-to-node through the graph) in an effort to cluster the $\upalpha$ colonies based on their likely parent $\upbeta$ grain lineage. We present the results of the flood fill clustering in Figure~\ref{subfig:betagraph}.

\begin{figure}[htbp!]
    \centering
    \includegraphics[height=0.325\textwidth]{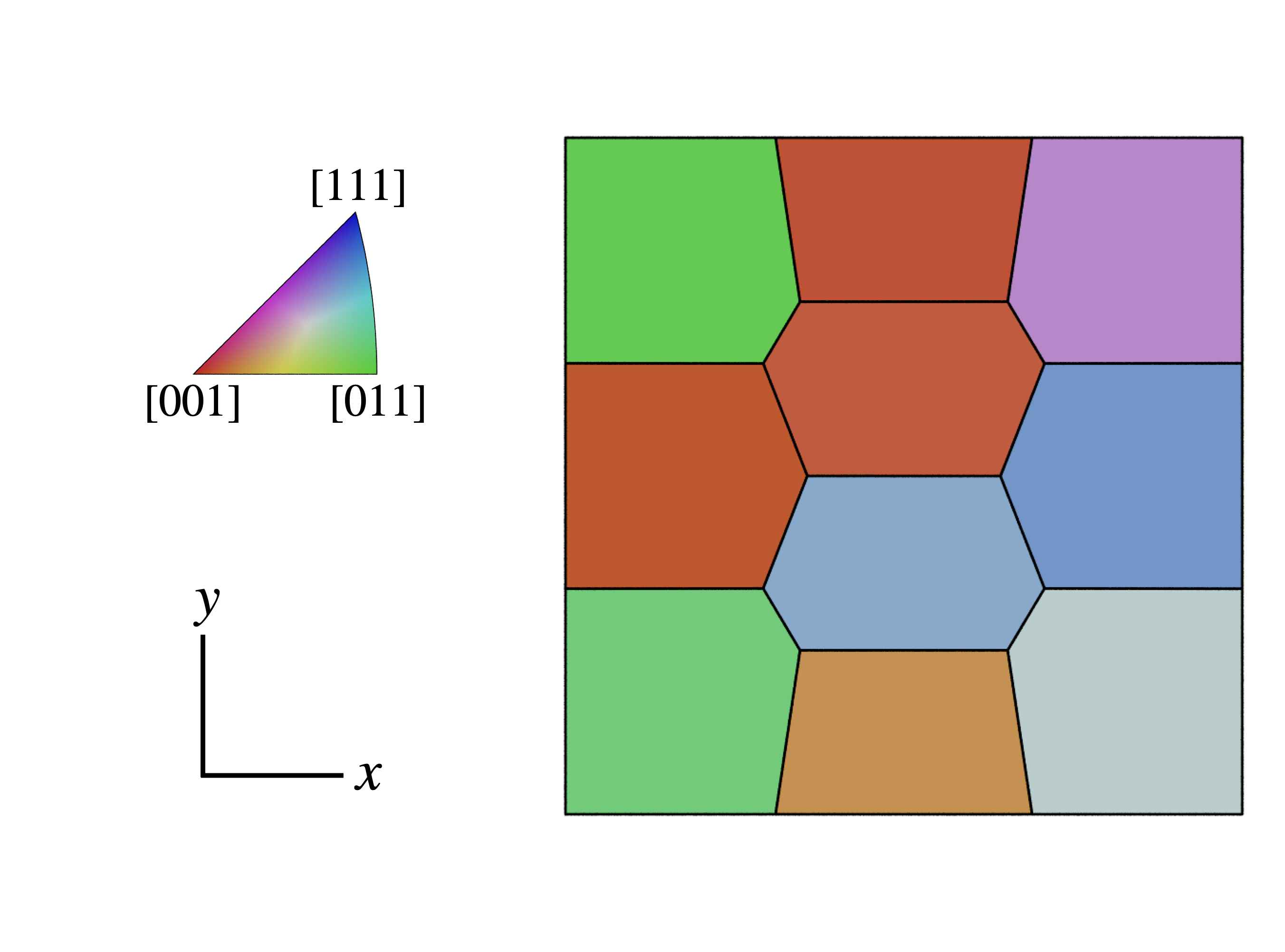}
    \caption{An idealized two-dimensional polycrystalline sample, representing the reflected parent $\upbeta$ grain orientations, as calculated from the clustered $\upalpha$ colony fibers. The inverse pole figure map is plotted with respect to the sample $z$ axis (i.e., out of page).}
    \label{fig:2dbeta_pp}
\end{figure}

\begin{figure}[htbp!]
    \centering
    \subfigure[]{%
	\includegraphics[height=0.325\textwidth]{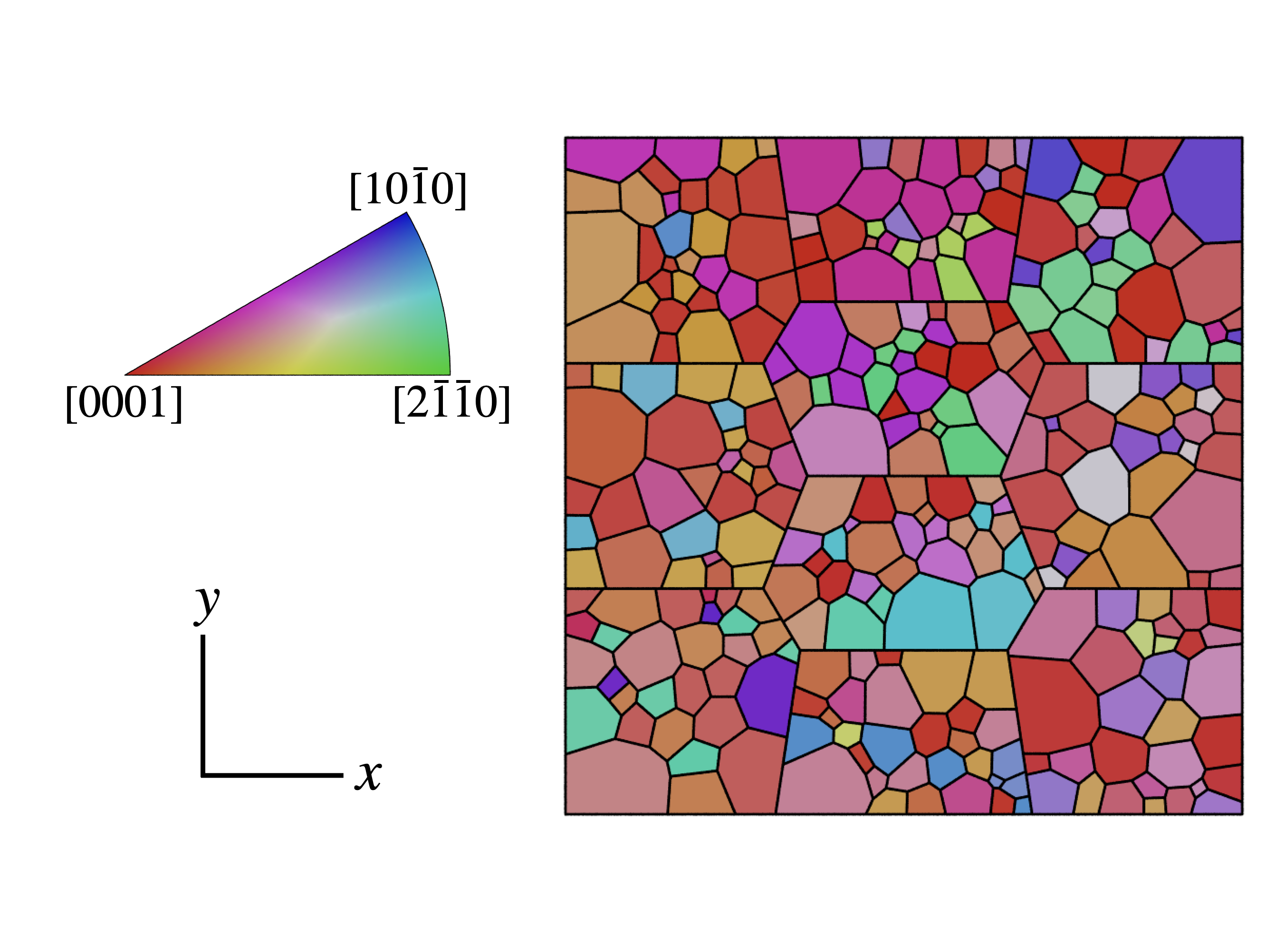}
        \label{subfig:2dalpha_vars1}}
    \subfigure[]{%
	\includegraphics[height=0.325\textwidth]{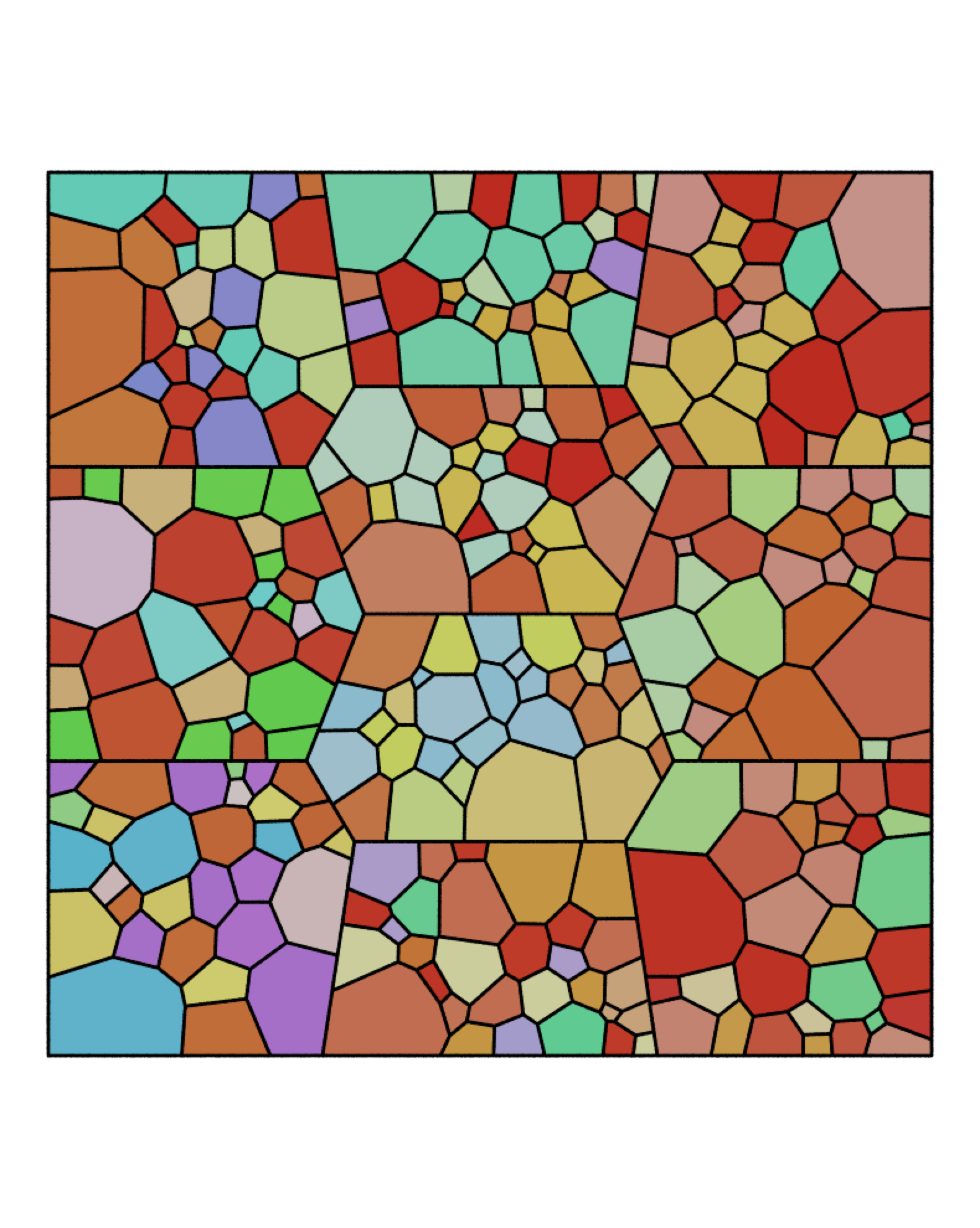}
        \label{subfig:2dalpha_vars2}}
    \subfigure[]{%
	\includegraphics[height=0.325\textwidth]{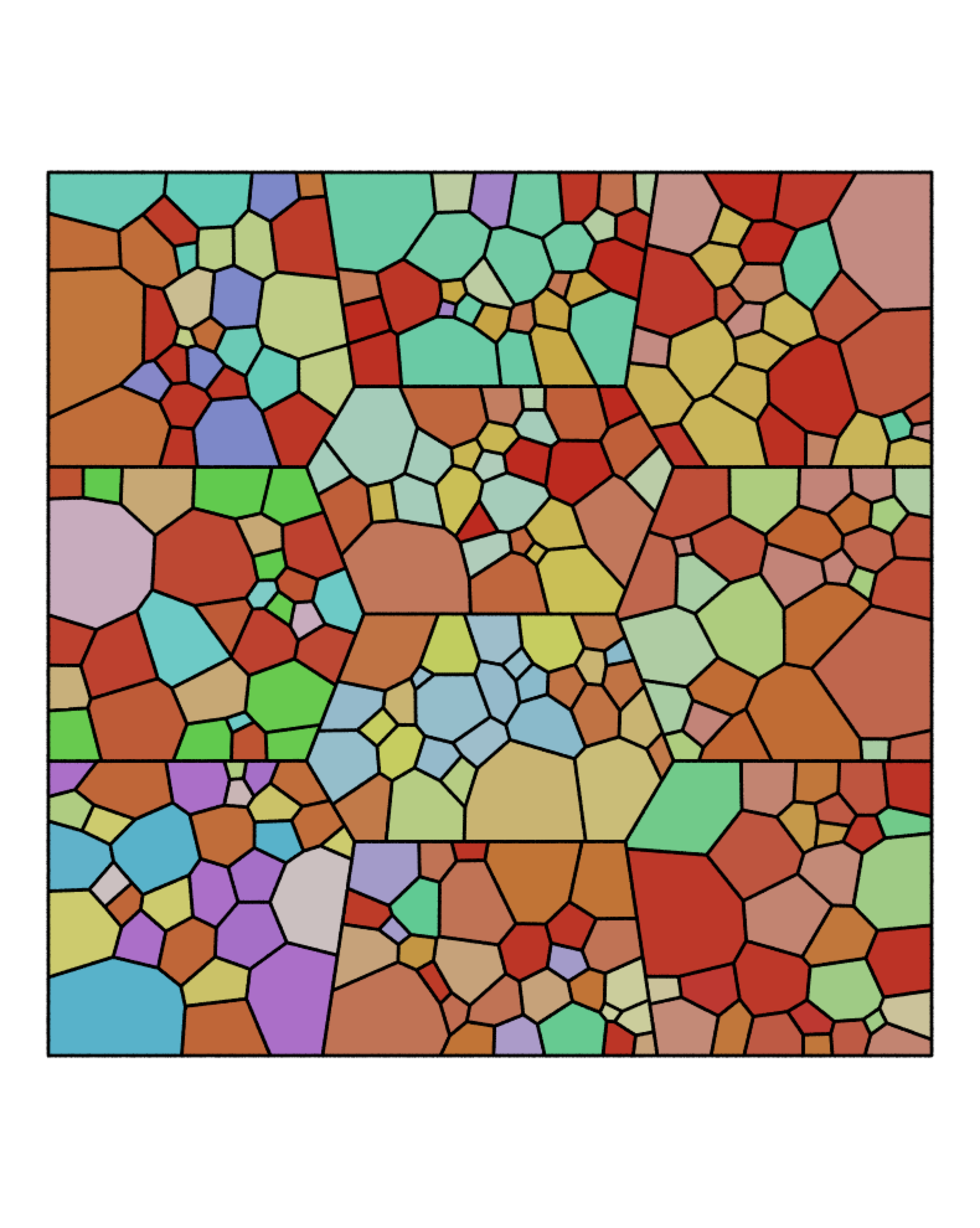}
        \label{subfig:2dalpha_vars3}}
    \caption{An idealized two-dimensional polycrystalline sample, representing~\subref{subfig:2dalpha_vars1} the paired orientations which share the same $c$ axes as the actual orientations,~\subref{subfig:2dalpha_vars2} those which are reflections of the actual orientations, and~\subref{subfig:2dalpha_vars3} those which are reflections of the paired orientations. The inverse pole figure map is plotted with respect to the sample $z$ axis (i.e., out of page).}
    \label{fig:2dalpha_vars}
\end{figure}

Next, we employ the method described in Section~\ref{subsubsec:registration} to calculate the two possible parent $\upbeta$ grain orientations for each of the sets of clustered $\upalpha$ colonies. We note that of the two sets back-calculated orientations, one of the sets is the true parent $\upbeta$ grain orientations (presented previously in Figure~\ref{subfig:2dbeta}). We further present the second set of reflected parent $\upbeta$ grain orientations that we would otherwise be unable to distinguish between in Figure~\ref{fig:2dbeta_pp}. We take care here to note that our method is unable to distinguish between the orientations presented in Figure~\ref{subfig:2dbeta} and Figure~\ref{fig:2dbeta_pp} \emph{independently per grain}---i.e., the reader should not misconstrue these figures as the only two possible fields of orientations, and indeed each grain each can take one of two possible orientations. From these two parent $\upbeta$ grain orientations, we finally calculate the 24 (total) $\upalpha$ variants (Section~\ref{subsubsec:forwardalpha}), and compare their $c$ axes against each $\upalpha$ colony's measured $c$ axis and its reflection. In total, four variants are possible for each measured $\upalpha$ fiber, and are presented in Figure~\ref{subfig:2dalpha} (i.e., the true orientation set), as well as the three other possible orientation sets in Figure~\ref{fig:2dalpha_vars}. We again take care here to note that our method is unable to distinguish between the orientations presented in Figure~\ref{subfig:2dalpha} and Figure~\ref{fig:2dalpha_vars} \emph{independently per colony}---i.e., the reader should not misconstrue these figures as the only four possible fields of orientations, and indeed each colony each can take one of four possible orientations. In total, we demonstrate that the fibers measured via PLM allow for the determination of the $\upalpha$ colony orientations to any of the possibilities presented in Figures~\ref{subfig:2dalpha} and~\ref{fig:2dalpha_vars}.

\section{Crystal Plasticity Simulations}
\label{sec:cpsims}

In a previous study~\cite{vanWees2023}, we demonstrated that the random choice of the third Euler-Bunge angle (i.e., based on what PLM was able to measure) had a potentially large effect on the predicted deformation response of crystal plasticity finite element (CPFEM) simulations. Specifically, we found that the elastic responses between samples in which the third Euler-Bunge angle was randomly chosen did not differ appreciably, as expected owing to the transversely isotropic nature of HCP crystals~\cite{bower,nye} and thus the negligible impact of rotation about the $c$ axis on purely elastic response. However, as plasticity developed, we observed that the deviations in stress predictions between samples grew considerably to average errors of approximately 20\%, to the point where PLM could not be effectively used as an instantiation method for CPFEM simulations in which the prediction past the elastic regime was of interest.

We use this previous study as inspiration to test the potential consequences of the quasi-deterministic approach to finding the $\upalpha$ colony orientations on the predicted deformation response of virtual samples. Here, we construct a virtual polycrystalline sample which exhibits a parent $\upbeta$ grain / $\upalpha$ colony morphology and assign $\upalpha$ colony orientations based on the Burgers orientation relationship and their shared parent $\upbeta$ grain lineage. We perform simulations on multiple variations of this baseline microstructure, where we randomly choose the orientations of the $\upalpha$ colonies between one of the four variants we calculate via our quasi-deterministic approach. We then provide systematic comparisons between these simulations in an effort to deduce the potential effect on deformation predictions, primarily by inspecting the prediction of stress fields in 3D.

We provide a brief description of the crystal plasticity framework utilized in this study in Appendix~\ref{sec:cpfem}. Otherwise, below we describe the construction of a suite of simulations probing the effects of the choice of quasi-deterministic orientations and their results.

\subsection{Virtual Sample Generation}
\label{subsubsec:virtualsample}

\subsubsection{Microstructure Morphology}

To generate idealized microstructures for use in simulations, we utilize multilevel tessellations (first described in~\cite{Kasemer2017} for similar use) as realized by the software package Neper~\cite{neper}. In multilevel tessellations, we may perform successive levels of discretization by considering the sub-volumes from the previous scale of tessellation as domains available for further discretization, overall in an effort to reach a desired morphology. Here, we aim to maintain the parent $\upbeta$ grain morphology with sub-grain discretization to create volumes representing $\upalpha$ colonies which collectively maintain their parent $\upbeta$ grain morphology.

\begin{figure}[htbp!]
    \centering
    \subfigure[]{%
	\includegraphics[height=0.325\textwidth]{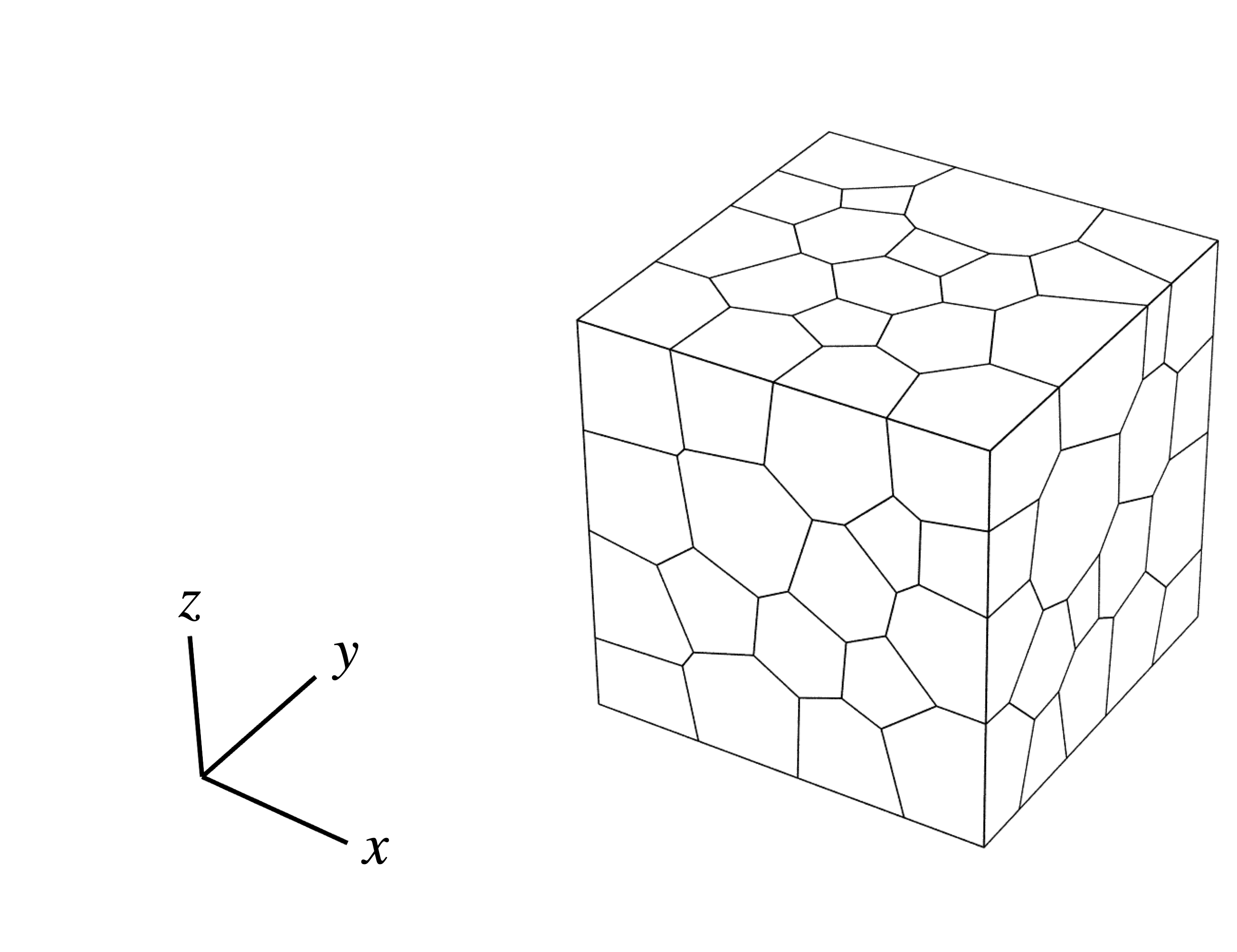}
        \label{subfig:betamorpho}}
    \subfigure[]{%
	\includegraphics[height=0.325\textwidth]{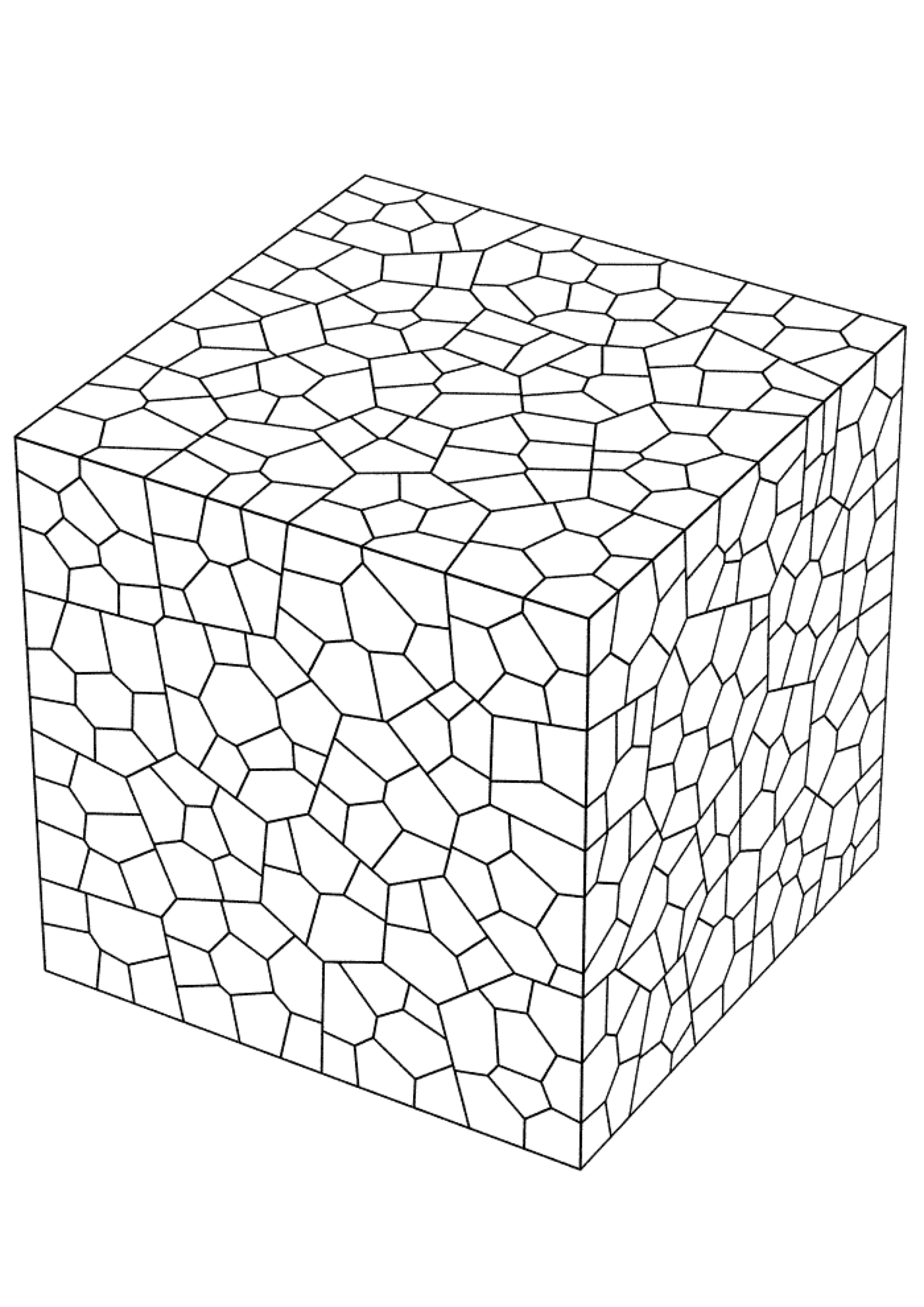}
        \label{subfig:alphamorpho}}
    \caption{\subref{subfig:betamorpho} First level tessellation depicting the morphology of the parent $\upbeta$ grains, and \subref{subfig:alphamorpho} second level tessellation depicting the morphology of the $\upalpha$ colonies that transformed from parent $\upbeta$ grains.}
    \label{fig:morpho}
\end{figure}

To achieve this, we perform a first level tessellation on a \qtyproduct[product-units=repeat]{1 x 1 x 1}{\milli\meter} domain. The first level of the tessellation contains 100 grains, which represent the morphology of the parent $\upbeta$ grains present at a temperature above the transus. As we model intend to model only the $\upalpha$ phase in the simulations, the generation of the parent $\upbeta$ grains is only to achieve the desired transformation microstructure, and the information of the parent $\upbeta$ grains are not explicitly retained for use in the simulations. We utilize Laguerre tessellations~\cite{Kasemer2017,Quey2018} to set target distributions for the normalized equivalent diameter and the sphericity to generate a highly equiaxed morphology~\cite{neperweb,neperfepx}. For the normalized equivalent diameter, we utilize a normal distribution with mean of 1 and standard deviation of 0.15, and for the sphericity (specifically $1-\hbox{sphericity}$), we utilize a lognormal distribution with mean of 0.145 and standard deviation of 0.03. We present the results of the first level of tessellation, representative of parent $\upbeta$ grains, in Figure~\ref{subfig:betamorpho}.

We next perform a second level of tessellation. Here, we consider each parent $\upbeta$ grain from the first level tessellation separately as a domain, and further sub-discretize each parent $\upbeta$ grain into a number of $\upalpha$ colonies. We utilize Laguerre tessellations for the second level tessellation, where we set target distributions for the grain diameter to an absolute mean of \SI{108.4}{\micro\meter} and a standard deviation of \SI{5.8}{\micro\meter}, and sphericity to the same as in the first level tessellations, which attempts to enforce a highly normalized structure for the $\upalpha$ colonies---i.e., we enforce that each $\upalpha$ colony in the domain has similar volume and shape---and target approximately 1500 total colonies (we note that due to variations in parent $\upbeta$ grain sizes, some parent grains may have more $\upalpha$ colonies than others). We present the results of the second level of tessellation, representative of $\upalpha$ colonies, in Figure~\ref{subfig:alphamorpho}, and note that the domain has 1503 total $\upalpha$ colonies (or approximately 15 $\upalpha$ colonies per parent $\upbeta$ grain on average). In this way, we maintain that the $\upalpha$ colonies that share the same parent $\upbeta$ grain lineage are spatially contiguous and collectively maintain the morphology of their parent $\upbeta$ grain. We finally generate a geometry-conforming finite element mesh, again via Neper, with a high enough resolution to produce consistent intra-grain results.

\subsubsection{Orientation Generation}
\label{subsubsec:origen}

With the geometry of the microstructure and an attendant finite element mesh generated, we now turn to the assignment of orientations. We again note that the simulations we perform in this study consider only the $\upalpha$ phase of Ti64, and ultimately the orientations which we assign to the grains/elements present in the virtual sample will be for the $\upalpha$ phase. To maintain the transformation microstructure and the attendant local $\upalpha$ texture that arises due to the $\upbeta$ to $\upalpha$ transformation, however, we begin first by generating a set of $\upbeta$ orientations assigned to the parent $\upbeta$ grains (i.e., the first level tessellation). For each $\upalpha$ colony within a given parent $\upbeta$ grain, we utilize the Burgers orientation relationship (Equation~\ref{eq:bor}) to calculate the 12 $\upalpha$ colony variants that may arise from the parent $\upbeta$ grain orientation. We then assign orientations to the $\upalpha$ colonies by randomly selecting from this group of 12 (i.e., we assume no variant preference). These $\upalpha$ orientations are what are assigned directly to the elements within an $\upalpha$ colony, and we assume that each $\upalpha$ colony is initially a single crystal (i.e., all elements in a single $\upalpha$ colony are assigned the same initial orientation).

\begin{figure}[htbp!]
    \centering
    \subfigure[]{%
	\includegraphics[height=0.325\textwidth]{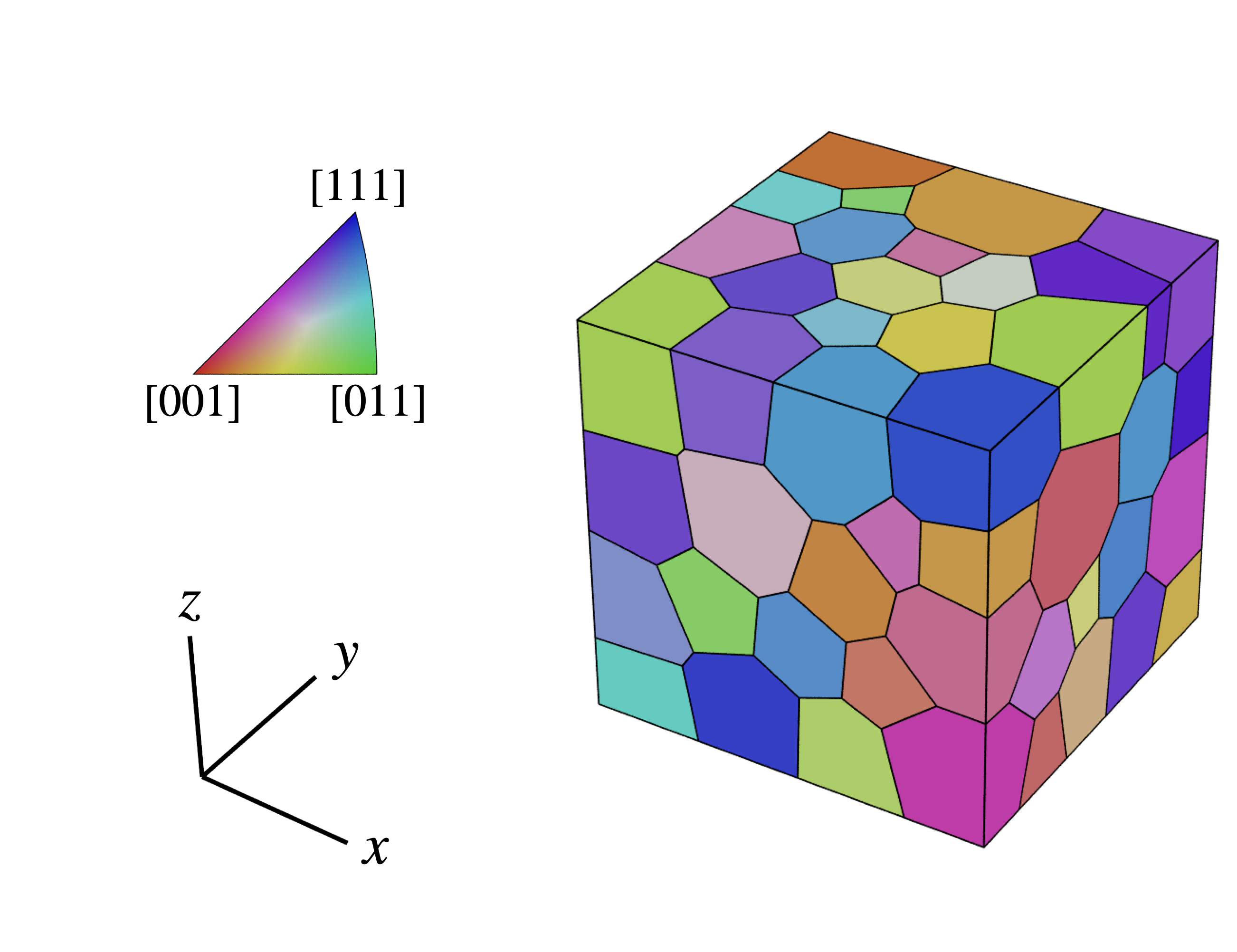}
        \label{subfig:betamorphooris_rand}}
    \subfigure[]{%
	\includegraphics[height=0.325\textwidth]{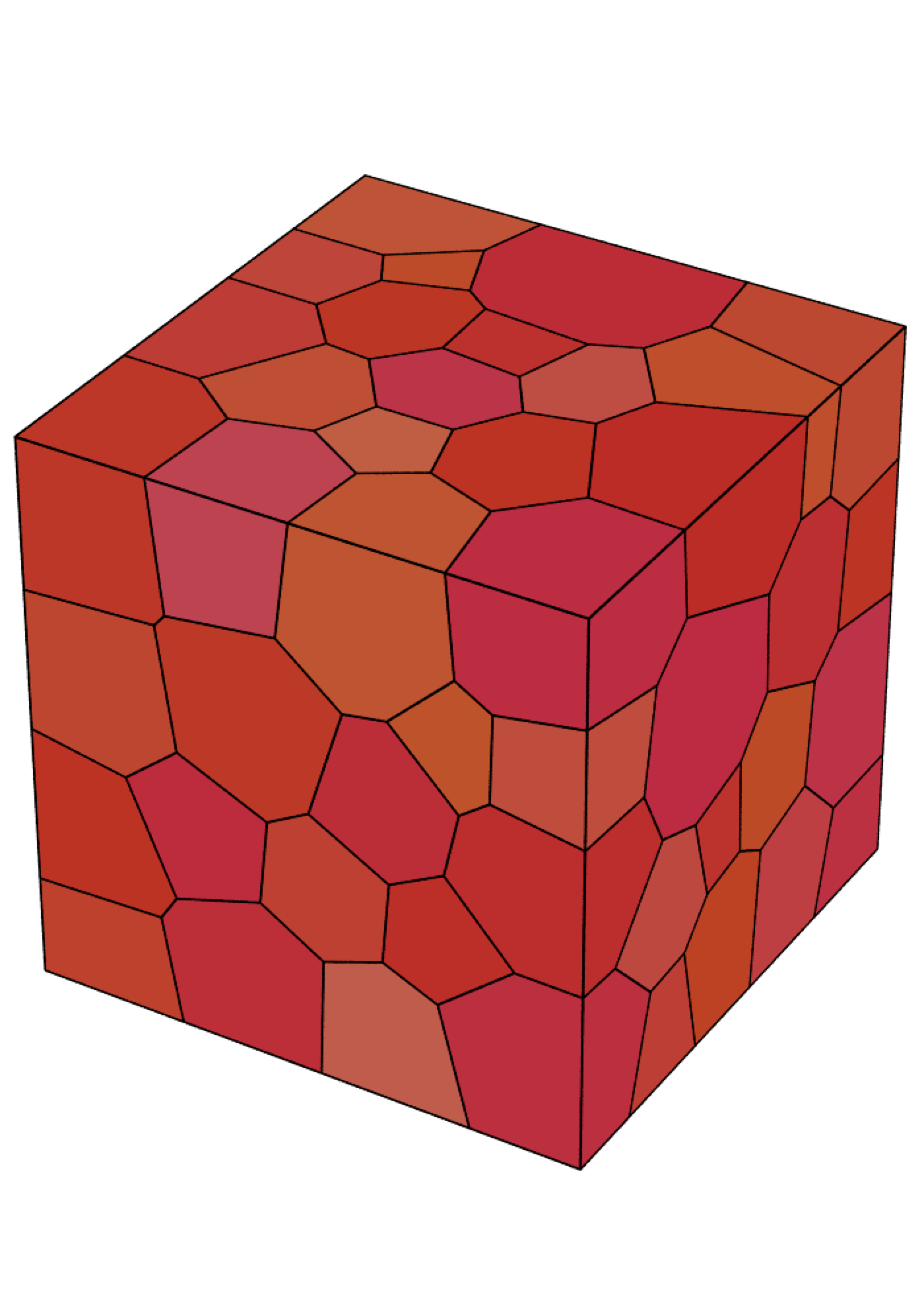}
        \label{subfig:betamorphooris_cube5}}
    \subfigure[]{%
	\includegraphics[height=0.325\textwidth]{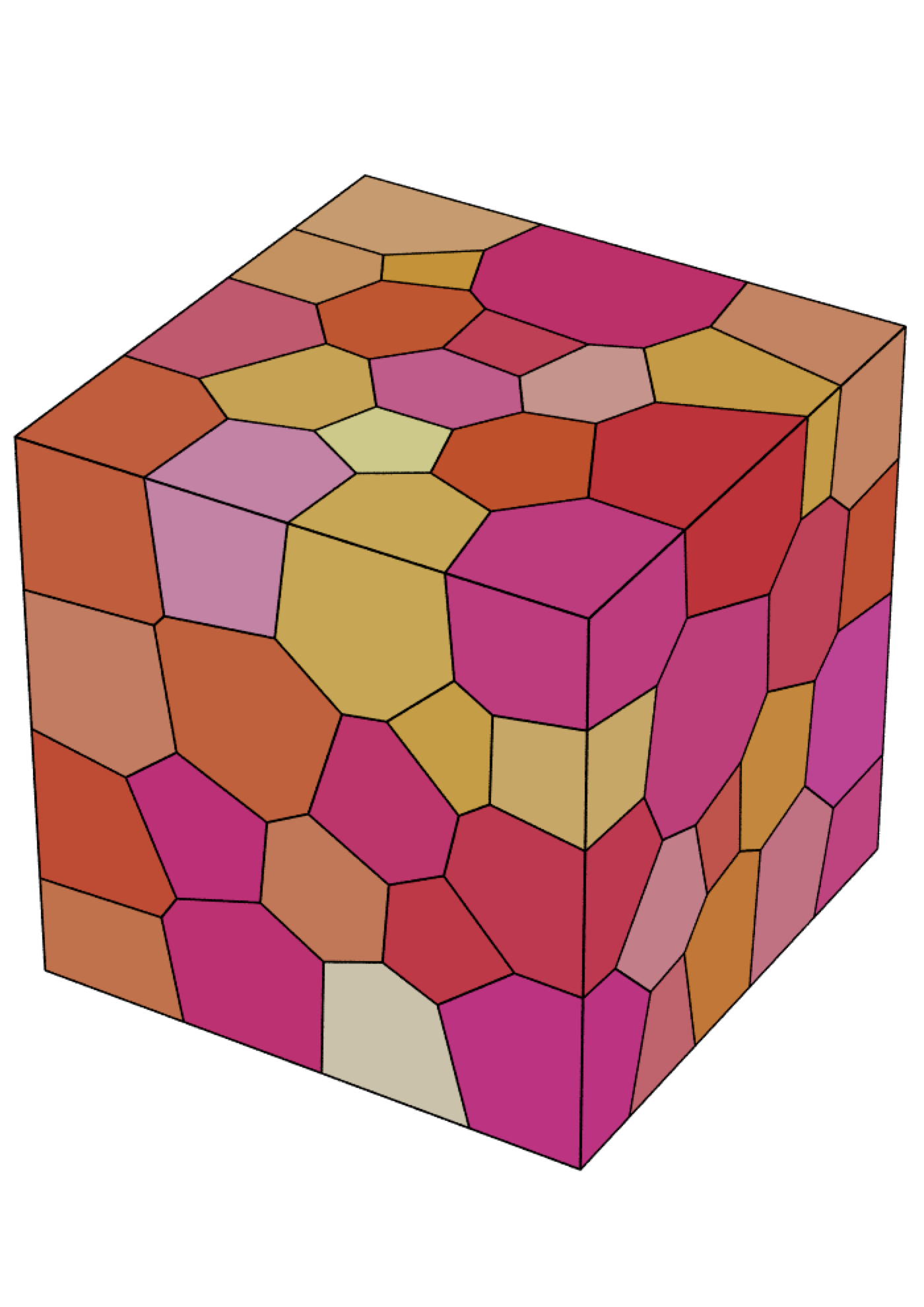}
        \label{subfig:betamorphooris_cube15}}
    \caption{The parent $\upbeta$ grains colored by orientation (via the IPF color map with respect to the sample $z$ direction) for the samples with orientations sampled from \subref{subfig:betamorphooris_rand} a random distribution, \subref{subfig:betamorphooris_cube5} a cube texture distribution with a average deviation of \SI{5}{\degree}, and \subref{subfig:betamorphooris_cube15} a cube texture distribution with an average deviation of \SI{15}{\degree}.}
    \label{fig:morphobetaoris}
\end{figure}

\begin{figure}[htbp!]
    \centering
    \subfigure[]{%
	\includegraphics[height=0.325\textwidth]{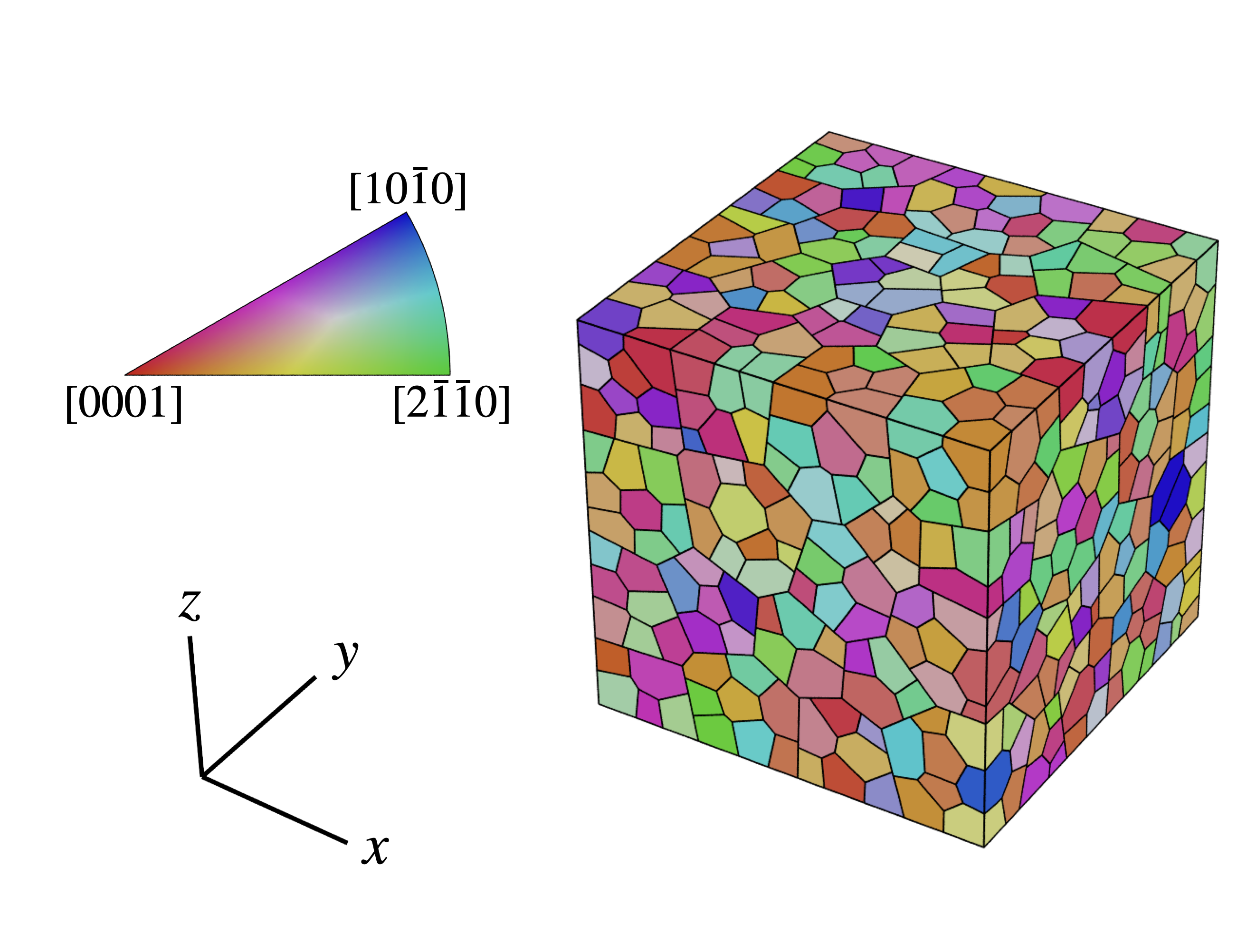}
        \label{subfig:alphamorphooris_rand}}
    \subfigure[]{%
	\includegraphics[height=0.325\textwidth]{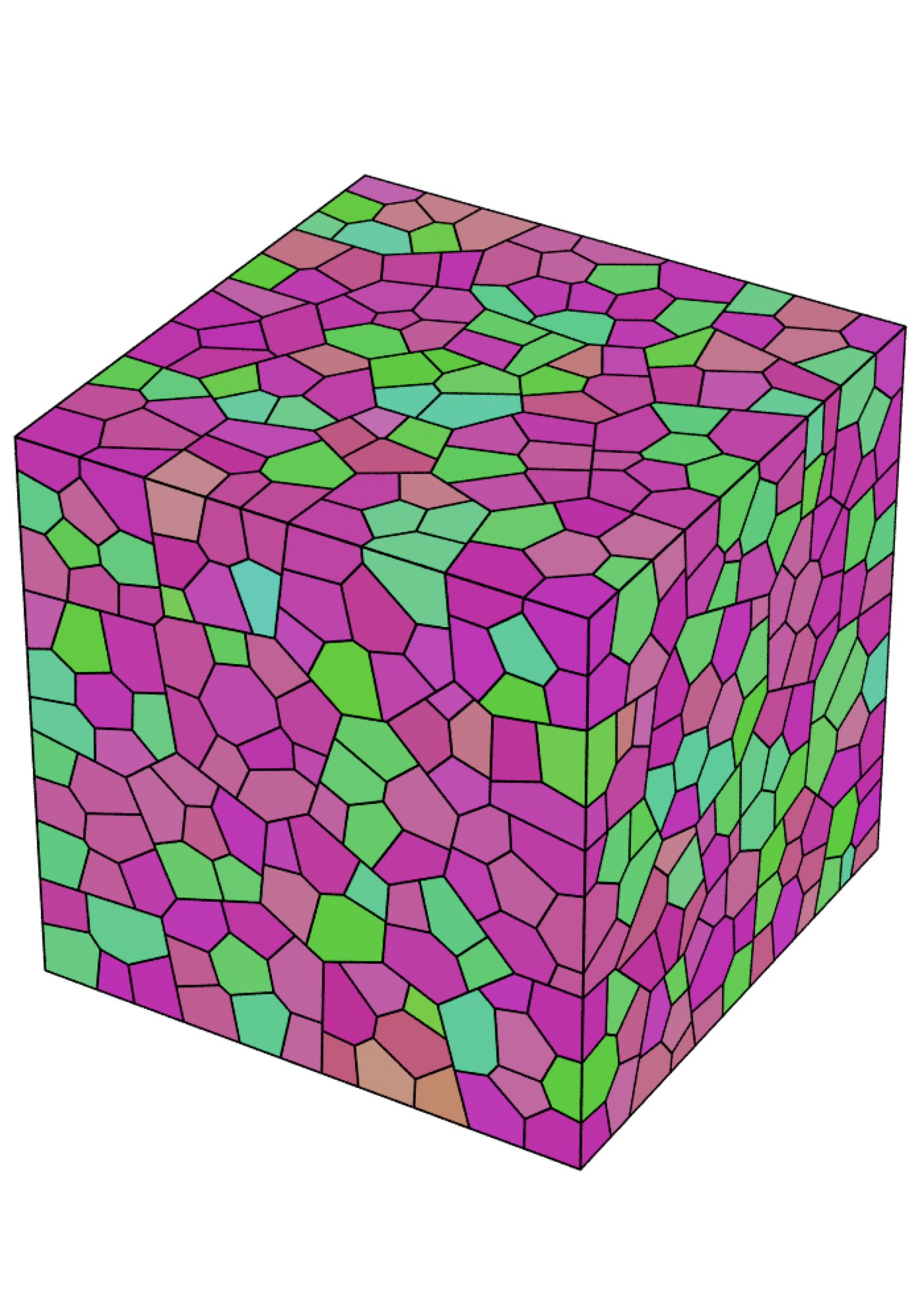}
        \label{subfig:alphamorphooris_cube5}}
    \subfigure[]{%
	\includegraphics[height=0.325\textwidth]{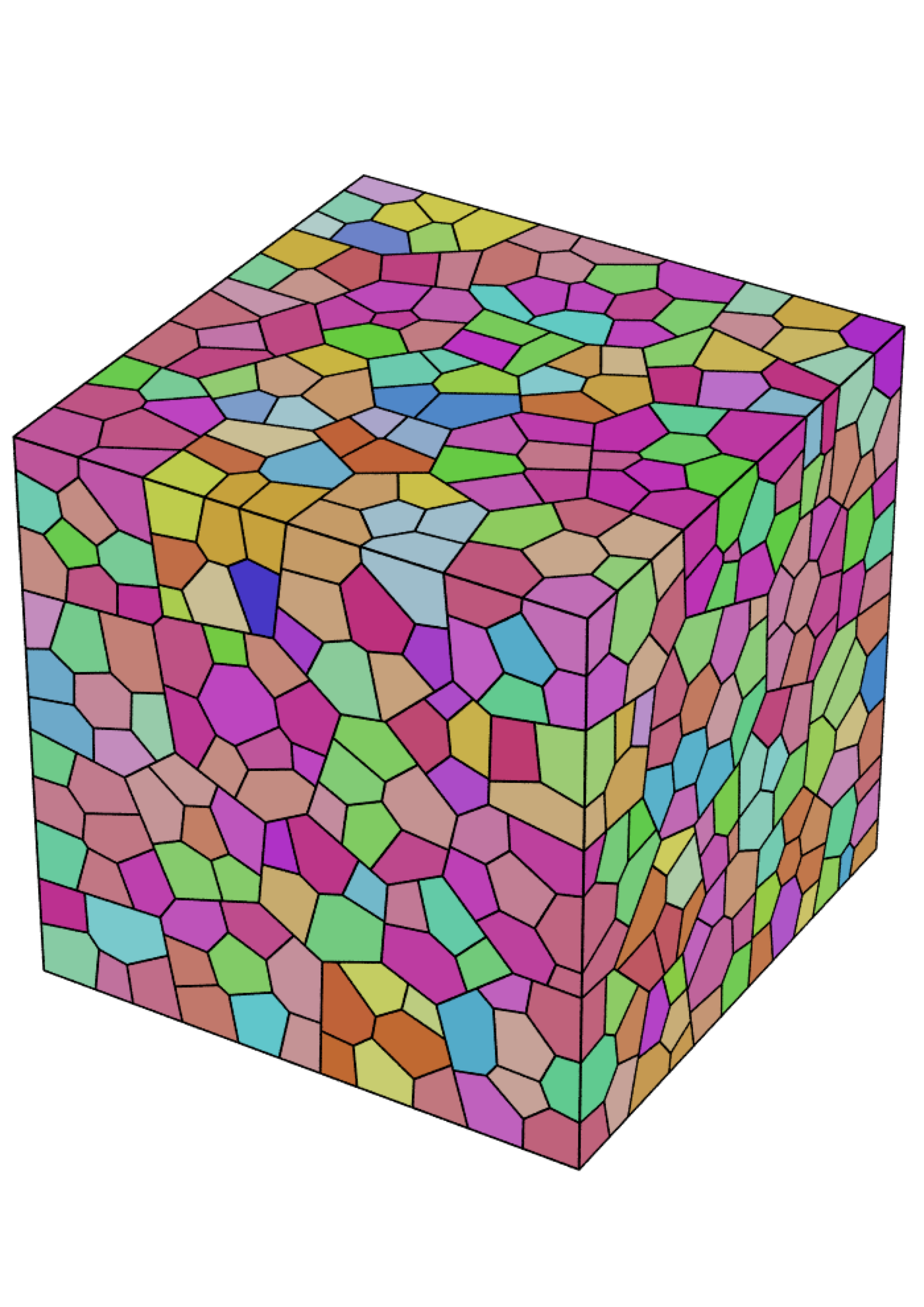}
        \label{subfig:alphamorphooris_cube15}}
    \caption{The $\upalpha$ colonies colored by orientation (via the IPF color map with respect to the sample $z$ direction) for the samples with parent $\upbeta$ grains orientations sampled from \subref{subfig:alphamorphooris_rand} a random distribution, \subref{subfig:alphamorphooris_cube5} a cube texture distribution with a average deviation of \SI{5}{\degree}, and \subref{subfig:alphamorphooris_cube15} a cube texture distribution with an average deviation of \SI{15}{\degree}.}
    \label{fig:morphoalphaoris}
\end{figure}

Regarding the selection of the parent $\upbeta$ grain orientations, specifically, we assume two primary paradigms. In the first, we generate a random set of parent $\upbeta$ grain orientations. In the second, we generate a set of parent $\upbeta$ grain orientations assuming a cube texture~\cite{raabe}, as Ti64 is often processed (rolled) above the transus temperature when it is entirely $\upbeta$ phase (BCC), and cubic materials tend toward a cube texture during rolling. We plot the parent $\upbeta$ grains with three different texture assumptions (random, and two different strengths of cube texture) in Figure~\ref{fig:morphobetaoris}. We then plot the ultimate $\upalpha$ phase microstructures that arise from these parent $\upbeta$ grains in Figure~\ref{fig:morphoalphaoris}.

\subsubsection{Orientation Variation}
\label{subsubsec:samplevar}

As described in Section~\ref{subsubsec:origen}, we ultimately have three base samples (as depicted in Figure~\ref{fig:morphoalphaoris}). To understand the effect that the quasi-determinsitic approach to identifying the $\upalpha$ colony orientations (described in Section~\ref{sec:methodology}) has on the predicted deformation response, we generate a set of permuted samples in which the $\upalpha$ colony orientations are randomly chosen among the four possibilities that each colony may be based on the quasi-deterministic reconstruction method described in this study. This mimics the degree to which we can determine the $\upalpha$ colony orientations via our quasi-deterministic reconstruction method. We generate 10 different samples for each of the three base microstructures, where each of the $\upalpha$ colony orientations are randomly chosen among their four possibilities (i.e., 30 permuted samples in total). The predicted deformation response of these permuted samples are compared against the predicted deformation response of the base samples (i.e., the true samples). Further, we will compare the results of the simulations described here to those from a previous study~\cite{vanWees2023} in which the third Euler-Bunge angle was chosen entirely at random to demonstrate the superior predictive capability when utilizing orientations from the reconstruction method.

\subsection{Material and Modeling Parameters}
\label{subsubsec:materialparms}

We utilize the elastic and plastic modeling parameters for the $\upalpha$ phase of Ti-6Al-4V (Ti64) in this study. We choose Ti64, as it is exhibits a transformation microstructure that adheres to the Burgers orientation relationship~\cite{Lutjering2003}. Further, Ti64 has enjoyed a significant degree of study~\cite{Kasemer2017,Kasemer2017b,Wielewski2017,Dawson2018}, and consequently we have a high degree of confidence in the elastic and plastic modeling parameters. We summarize the crystal plasticity modeling parameters in Tables~\ref{tab:elas_properties} and~\ref{tab:plas_properties}. We do not consider the retained $\upbeta$ phase morphology in this study, as it represents a small volume fraction of the material at room temperature (less than 10\% total volume), is often segregated to grain boundaries (as in the case of the microstructure realized by the mill annealed processing route~\cite{Lutjering2003,Kasemer2017}), and due to the relatively high degree of uncertainty in its modeling parameters. Consequently, we construct our virtual samples considering only the $\upalpha$ phase (i.e., single phase simulations).
\begin{table}[htbp!]
    \centering
    \begin{tabular}{c c c c}
        {\bf $C_{11}$ (\SI{}{\giga\pascal})} & {\bf $C_{12}$ (\SI{}{\giga\pascal})} & {\bf $C_{13}$ (\SI{}{\giga\pascal})} & {\bf $C_{44}$ (\SI{}{\giga\pascal})} \\
        \hline
        169.7 & 88.7 & 61.7 & 42.5 \\
    \end{tabular}
    \caption{Single crystal elastic constants for the $\upalpha$ phase of Ti-6Al-4V~\cite{Kasemer2017}.}
    \label{tab:elas_properties}
\end{table}
\begin{table}[htbp!]
    \centering
    \begin{tabular}{c c c c c c c}
        {\bf $\tau_{0,b}$ (\SI{}{\mega\pascal})} & {\bf $\tau_{0,p}$ (\SI{}{\mega\pascal})} & {\bf $\tau_{0,\pi}$ (\SI{}{\mega\pascal})} & {\bf $m$} & {\bf $\dot{\gamma_{0}}$ (\SI{}{s^{-1}})} & {\bf $h_{0}$ (\SI{}{\mega\pascal})} & {\bf $\tau_{s}$ (\SI{}{\mega\pascal})}  \\
        \hline
        390 & 468 & 663 & 0.01 & 1 & 190 & 530 \\
    \end{tabular}
    \caption{Plasticity modeling parameters for the $\upalpha$ phase of Ti-6Al-4V~\cite{Kasemer2017}.}
    \label{tab:plas_properties}
\end{table}

\subsection{Boundary Conditions and Load History}
\label{subsubsec:bcsandload}

We deform the specimens in uniaxial tension at a rate of \SI{1e-3}{} (i.e., quasi-static) in the sample $z$ direction using minimal boundary conditions~\cite{fepxweb} that minimize boundary conditions effects not representative of uniaxial tension. We apply deformation through 50 equally spaced time steps to a strain of \SI{0.05}{} (i.e., 50 time steps of \SI{1}{\second}), which allows for the close inspection of various points along the loading history.

\subsection{Simulation Results}
\label{subsec:simresults}

Here, we present the results of the simulations. To gain a broad understanding of the effects of the choice of orientations on the predictions of the deformation field, we compare the equivalent stress fields between the three base sample and their permutations with randomly varied orientations. We borrow the metric utilized in~\cite{vanWees2023} to quantify the error in stress predictions element-to-element between a single permuted sample and the base sample:
\begin{equation}
    \delta_{elt} = \frac{\lvert \sigma_b - \sigma_p \rvert}{\tilde{\sigma}_b} \cdot 100 \quad,
\end{equation}
where $\sigma_b$ is the equivalent stress for an element in the base sample, $\sigma_p$ is the equivalent stress for the same element in the permuted sample, and $\tilde{\sigma}_b$ is the domain-averaged equivalent stress in the base sample.

\begin{figure}[htbp!]
    \centering
    \subfigure[]{%
	\includegraphics[width=2.05in]{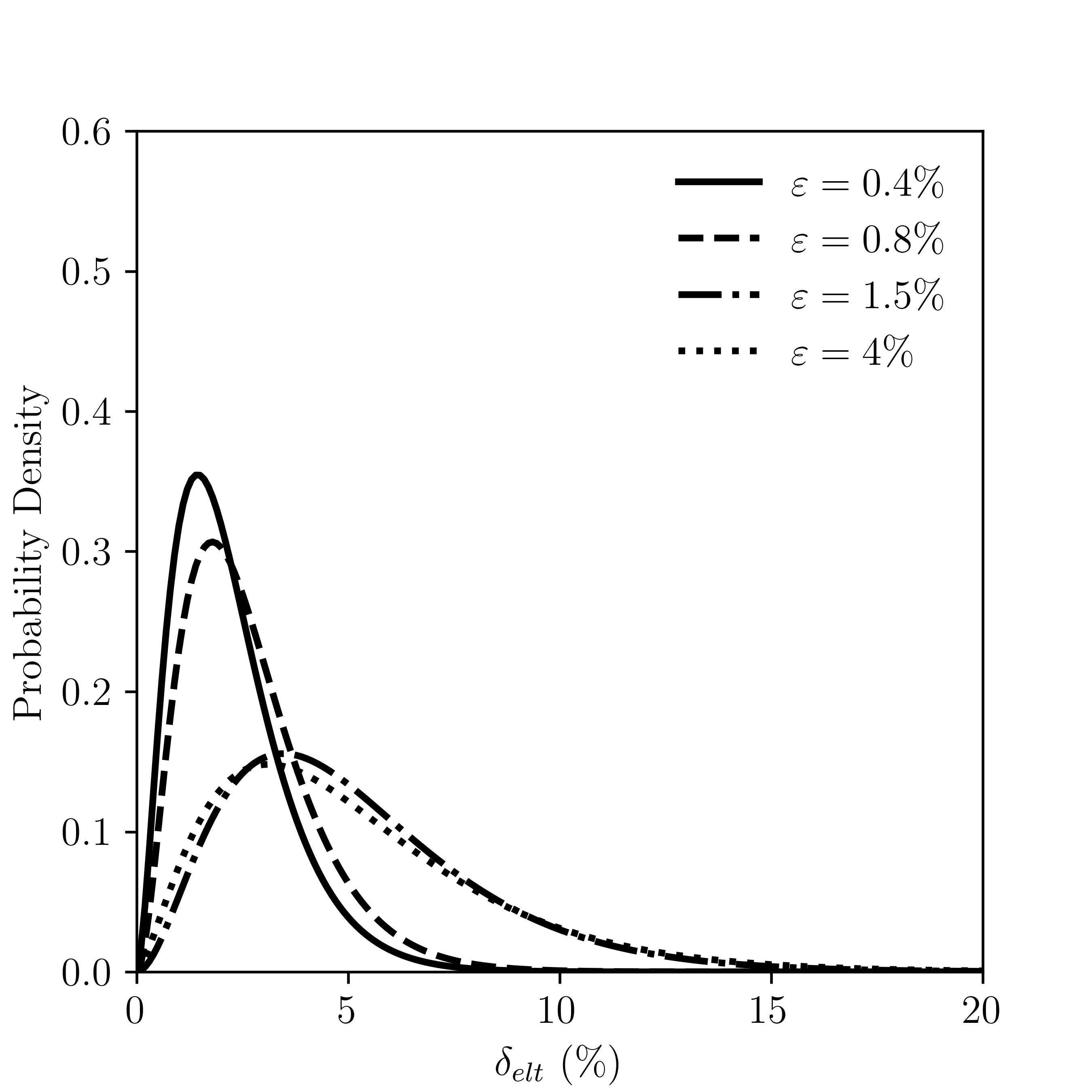}
        \label{subfig:resrand}}
    \subfigure[]{%
	\includegraphics[width=2.05in]{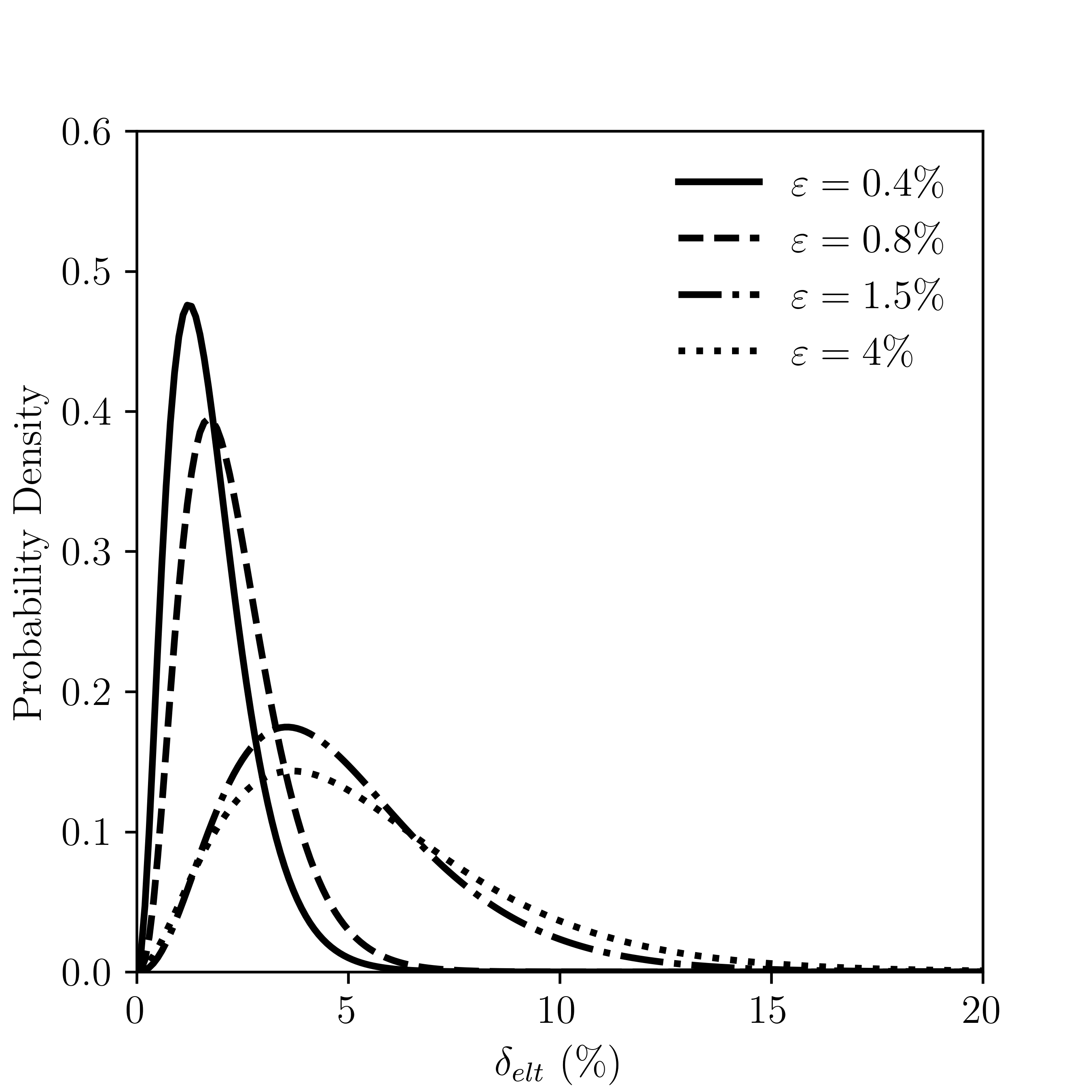}
        \label{subfig:restheta5}}
    \subfigure[]{%
	\includegraphics[width=2.05in]{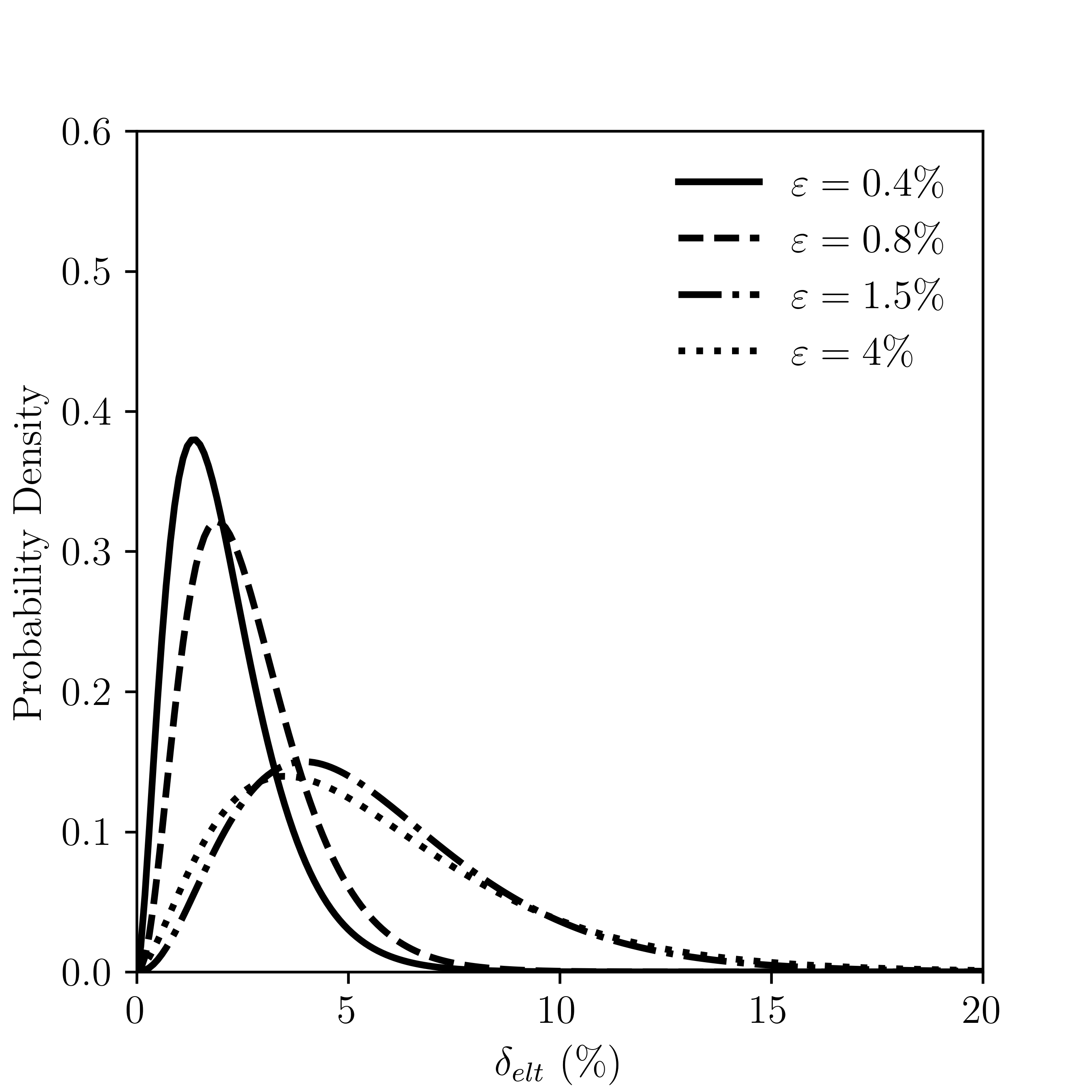}
        \label{subfig:restheta15}}
    \caption{Results depicting the average deviation of the elemental stresses predicted in the permuted samples against that predicted in the base samples at four different macroscopic strain states for the microstructures constructed with parent $\upbeta$ grain orientations generated from~\subref{subfig:resrand} a random texture,~\subref{subfig:restheta5} a cube texture with $\theta=\SI{5}{\degree}$ average misorientation, and~\subref{subfig:restheta15} a cube texture with $\theta=\SI{15}{\degree}$ average misorientation.}
    \label{fig:simresults}
\end{figure}

We simulate the deformation response of 10 permuted samples for each of the three base microstructures. We next average the $\delta_{elt}$ values for each element over the 10 different samples. This provides a statistical understanding of the effects of the choice of variant selection on the predicted deformation fields. We perform this calculation at four different macroscopic strain states of $\epsilon =$ 0.4\% (in the elastic regime), 0.8\% (shortly prior to macroscopic yield), 1.5\% (shortly after macroscopic yield), and 4.0\% (in the fully-developed plasticity regime). We plot these results in Figure~\ref{fig:simresults}, with statistics presented in Table~\ref{tab:simresults}.
\begin{table}[H]
    \centering
    \begin{tabular}{c|c c c c}
        Sample & $\epsilon=0.4\%$ & $\epsilon=0.8\%$ & $\epsilon=1.5\%$ & $\epsilon=4.0\%$ \\
        \hline
        Random & 2.24\% (1.34\%) & 2.66\% (1.52\%) & 5.21\% (3.01\%) & 5.19\% (3.29\%) \\
        Cube $\theta=\SI{5}{\degree}$ & 1.77\% (0.97\%) & 2.27\% (1.13\%) & 4.93\% (2.60\%) & 5.61\% (3.29\%) \\
        Cube $\theta=\SI{15}{\degree}$ & 2.10\% (1.25\%) & 2.66\% (1.42\%) & 5.63\% (3.06\%) & 5.62\% (3.43\%)
    \end{tabular}
    \caption{Statistics on the distributions of average deviation in the elemental stress predictions between the base sample and the permuted samples, specifically delimited as the mean and standard deviation (in parentheses).}
    \label{tab:simresults}
\end{table}

We observe that, generally, the results of the permuted samples compare more favorably to the base samples in the elastic regime compared to in the plastic regime. At macroscopic strain states of $\epsilon=0.4\%$ and $\epsilon=0.8\%$ (the latter shortly before macroscopic yield), the average errors across all three samples are 2.04\% and 2.53\%, respectively, while at macroscopic strain states of $\epsilon=1.5\%$ and $\epsilon=4.0\%$ (i.e., after macroscopic yield), the average errors across are three samples are 5.26\% and 5.47\%, respectively. The lower errors in the nominally elastic regime are consistent with the elastic transverse isotropic nature of HCP crystals~\cite{bower}, where we would expect rotations about the $c$ axes of crystals to have limited effect on the predicted stress field (rotations of the $c$ axes due to reflections, however, could have stronger effects). As plasticity develops, the differences in slip system activity will lead to more pronounced differences in the predicted deformation fields.

We next compare to our previous study~\cite{vanWees2023} in which the third Euler-Bunge angle was chosen entirely at random without employing the quasi-deterministic reconstruction method presented here. While results in the elastic regime between the two studies are broadly comparable (again, a consequence of the elastic transverse isotropic nature of HCP crystals), the results in the plastic regime differ greatly. Specifically, when utilizing a random choice of the four possible orientations as we have described in this study, the permuted simulations exhibit a significant reduction in error compared to when the third Euler-Bunge angle is chosen entirely at random. At $\epsilon=4.0\%$, for example, the average error in stress fields compared to the base samples across all three microstructures when using our quasi-deterministic reconstruction method is 5.47\%, compared to average deviations of approximately 23\% in our previous study when using entirely randomly chosen values for the third Euler-Bunge angle. In other words, randomly choosing one of the four quasi-deterministic orientations results in a more than four-fold reduction in deviation of stress field predictions from truth compared to an entirely random choice of the third Euler-Bunge angle. This large reduction in uncertainty lends significant credibility to PLM as a characterization method for informing CPFEM simulations of titanium alloys.

\section{Conclusion}
\label{sec:conclusion}

In this study, we have presented a method to reconstruct quasi-deterministic orientations for $\upalpha$ titanium measured via polarize light microscopy. Our method exploits the Burgers orientation relationship to back-calculate the likely orientations of the high temperature parent $\upbeta$ grains, from which the Burgers orientation relationship can be used to forward-calculate the deterministic room-temperature $\upalpha$ colony orientations. We can match these ideal variants against the PLM-measured fibers, and show that based on the ambiguities present in PLM, we can narrow down the $\upalpha$ orientations to four possibilities. We demonstrated our reconstruction method on a synthetic sample containing fully-known orientations for both the room temperature $\upalpha$ colony field as well as the parent $\upbeta$ grain field, which provides validation for our method's ability to calculate the quasi-deterministic orientations. 

We further demonstrated the effects of this quasi-deterministic approach when utilized to instantiate polycrystalline specimens for use in crystal plasticity finite element deformation simulations. Simulation results indicate that random selection of $\upalpha$ orientations based on their four possibilities results in significantly lower error in predicted stress response (on the order of 5\% or below) than when compared to simulation results when the rotation about the $c$ axis is entirely randomly chosen (upwards of 23\%). This indicates that while we cannot unambiguously determine the true orientation, the use of the quasi-deterministic orientations may be effectively employed to arrive at plausible predictions of the deformation field.

\section*{Acknowledgments}
\label{sec:acknowledgments}

We thank Karthik Shankar (University of Alabama) for assistance in the software development. We further thank Dr. Romain Quey for the development of Neper. AS was funded by the National Science Foundation through award DMR-2143808, and further utilized resources at The University of Alabama. MO acknowledges funding from the Air Force Research Laboratory. MK received support and resources from The University of Alabama.

\begin{appendices}
\setcounter{figure}{0}
\renewcommand\thefigure{A.\arabic{figure}}
\addcontentsline{toc}{section}{Appendix}

\section{Demonstration of Reconstruction Method in Rodrigues Space}
\label{sec:methodologyrod}

\begin{figure}[htbp!]
    \centering
    \subfigure[]{%
	\includegraphics[width=0.35\textwidth]{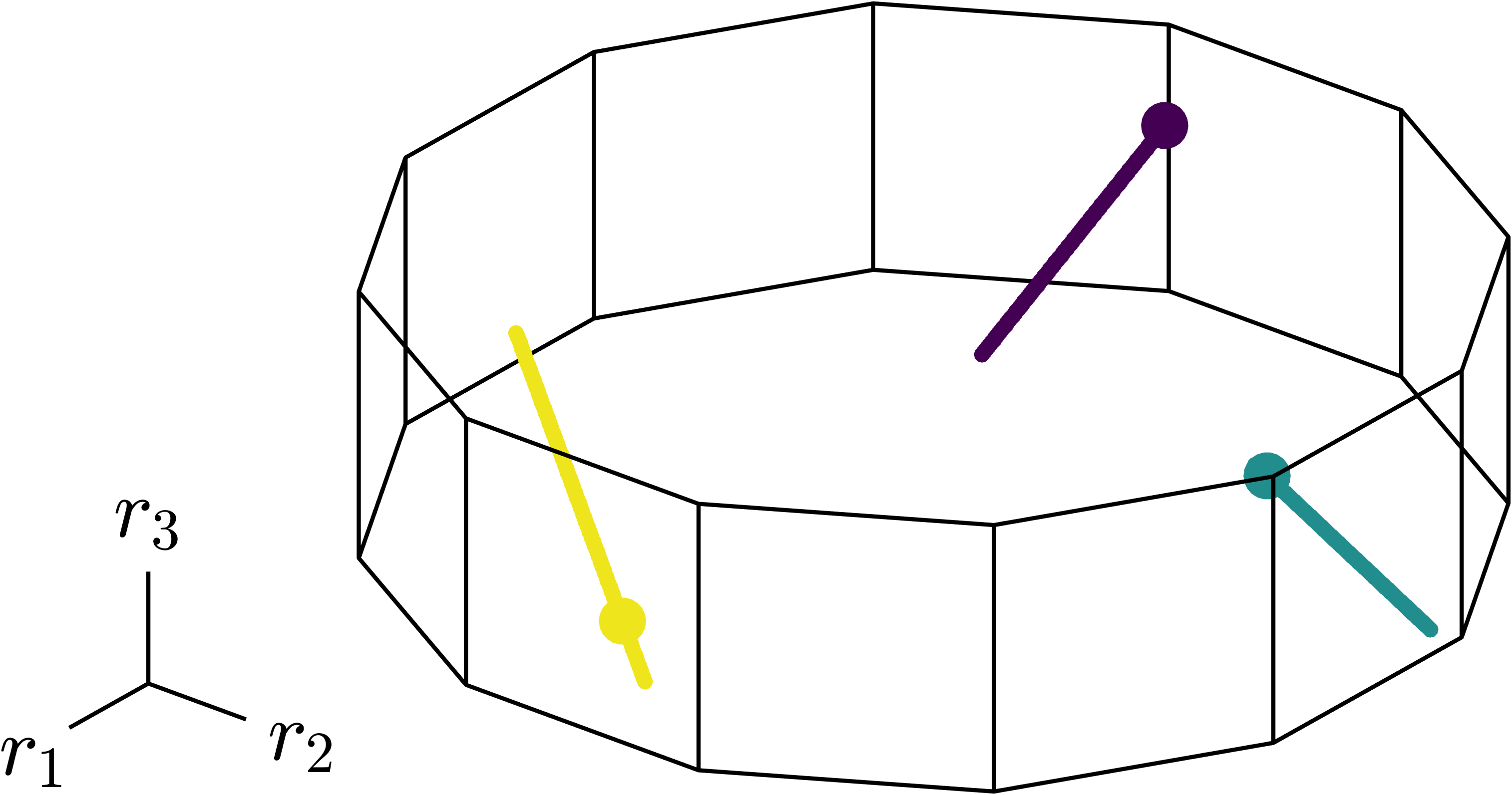}
        \label{subfig:alphaex}}
    \subfigure[]{%
	\includegraphics[width=0.35\textwidth]{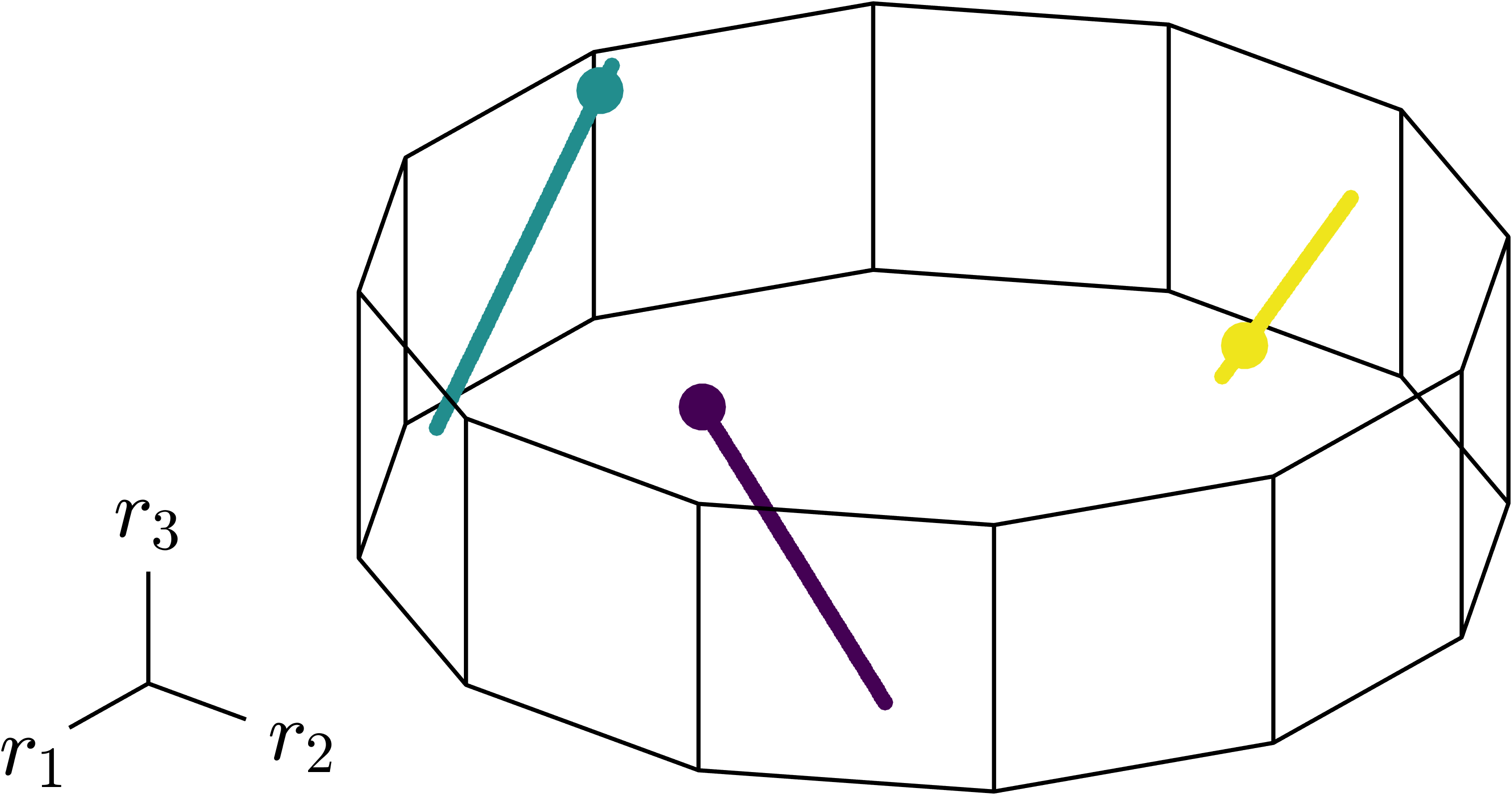}
        \label{subfig:alphaex2}}
    \caption{\subref{subfig:alphaex} Three random orientations (dots) plotted in the hexagonal symmetry fundamental region of Rodrigues space, and the corresponding crystallographic fibers depicting all orientations with the same $c$ axis orientations as their respective deterministic orientations (lines), and \subref{subfig:alphaex2} the reflected orientations and fibers. Note that the colors between the two subfigures are corresponding---i.e., the true orientations/fibers are colored the same as their reflected orientations/fibers.}
    \label{fig:alphaex}
\end{figure}

Here, we provide a demonstration of our reconstruction method in Rodrigues orientation space. The Rodrigues orientation parameterization~\citep{frank} is advantageous as it allows for the representation of all unique orientations via a relatively simple closed fundamental region, and due to the fact that crystallographic fibers present as straight lines in Rodrigues space~\citep{kumar,Quey2015}. To demonstrate this property, we present a visualization of three random orientations plotted in the hexagonal symmetry fundamental region of Rodrigues space in Figure~\ref{fig:alphaex}(a), along with their respective fibers of orientations who share the same $c$ axis orientations plotted as lines. PLM alone is unable to naturally distinguish between any of the orientations that lie on each line, since all orientations on each line share the same $c$ axis orientation (i.e., they are $c$ axis similar). Thus, this figure demonstrates the consequences of PLM's first limitation. We note that due to crystal symmetry, a fiber may exit one side of the fundamental region and appear on the opposite side, and thus all orientations which share the same $c$ axis may not be on a single line (though are collectively of the same fiber)~\citep{kumar}. Further, we plot the reflected orientations and fibers in Figure~\ref{fig:alphaex}(b).

\subsection{Demonstration Initialization}
\label{subsec:demonstration}

\begin{figure}[htbp!]
    \centering
    \includegraphics[width=0.3\textwidth]{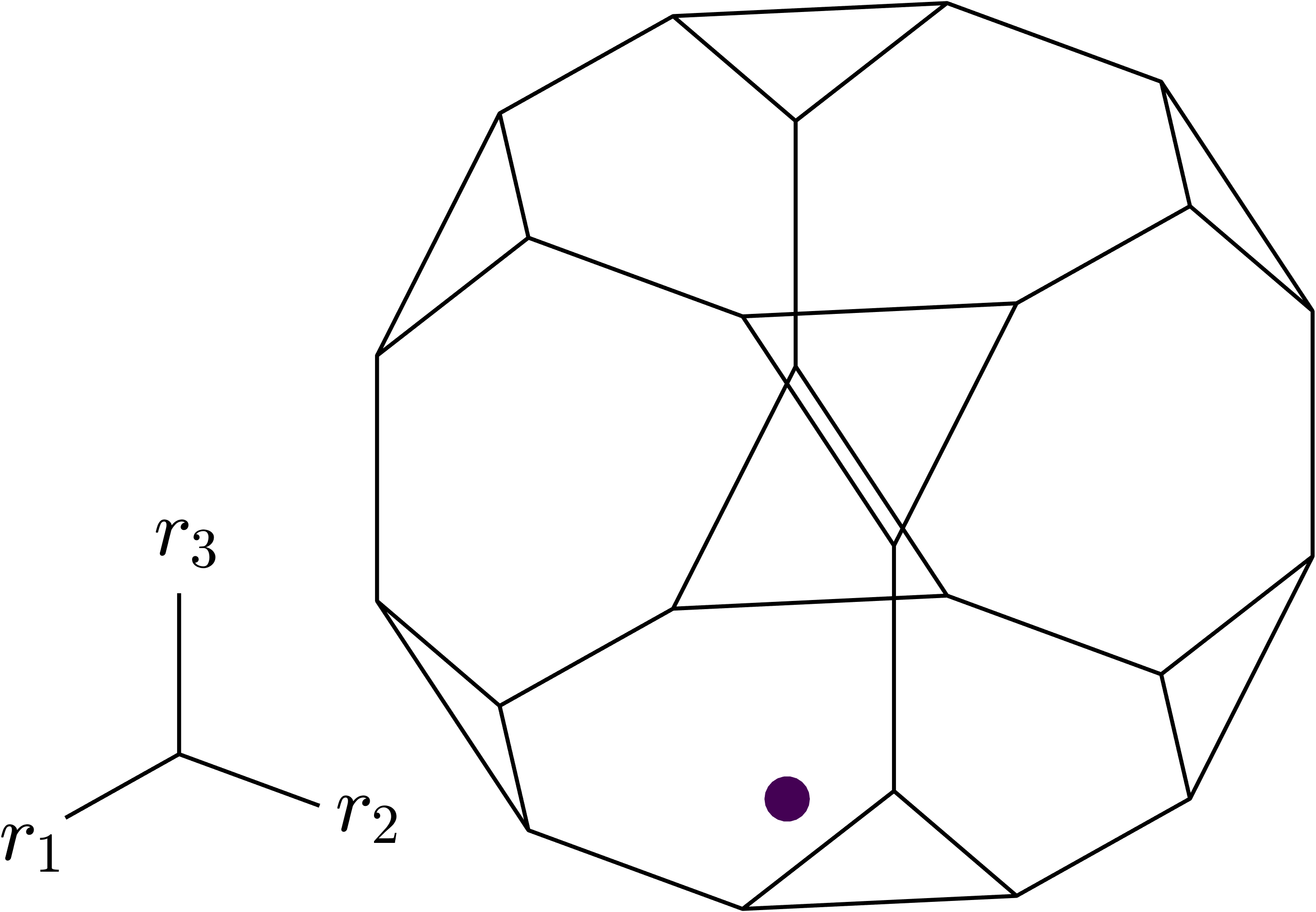}
    \caption{A random $\upbeta$ orientation plotted in the cubic symmetry fundamental region of Rodrigues space.}
    \label{fig:betaknown}
\end{figure}

To begin, we first generate a random parent $\upbeta$ grain orientation that represents the orientation of a grain which existed above the transus temperature. We plot this orientation in the cubic symmetry fundamental region of Rodrigues space in Figure~\ref{fig:betaknown}.

From this orientation, we calculate the orientations of the 12 $\upalpha$ colony variants that may arise due to the allotropic phase transformation. For these 12 $\upalpha$ colony variants, we further calculate the crystallographic fibers which represent the collection of crystallographic orientations with similar $c$ axis orientations to the $\upalpha$ colony variants (i.e., the consequence of the first orientation ambiguity described in Section~\ref{subsubsec:caxisambiguity}). For each of the 12 variant orientations/fibers, we calculate their reflected orientations/fibers by introducing a \SI{180}{\degree} rotation in $\phi_1$ (i.e., the consequence of the second orientation ambiguity described in Section~\ref{subsubsec:caxisambiguity2}).

\begin{figure}[htbp!]
    \centering
    \subfigure[]{%
	\includegraphics[width=0.35\textwidth]{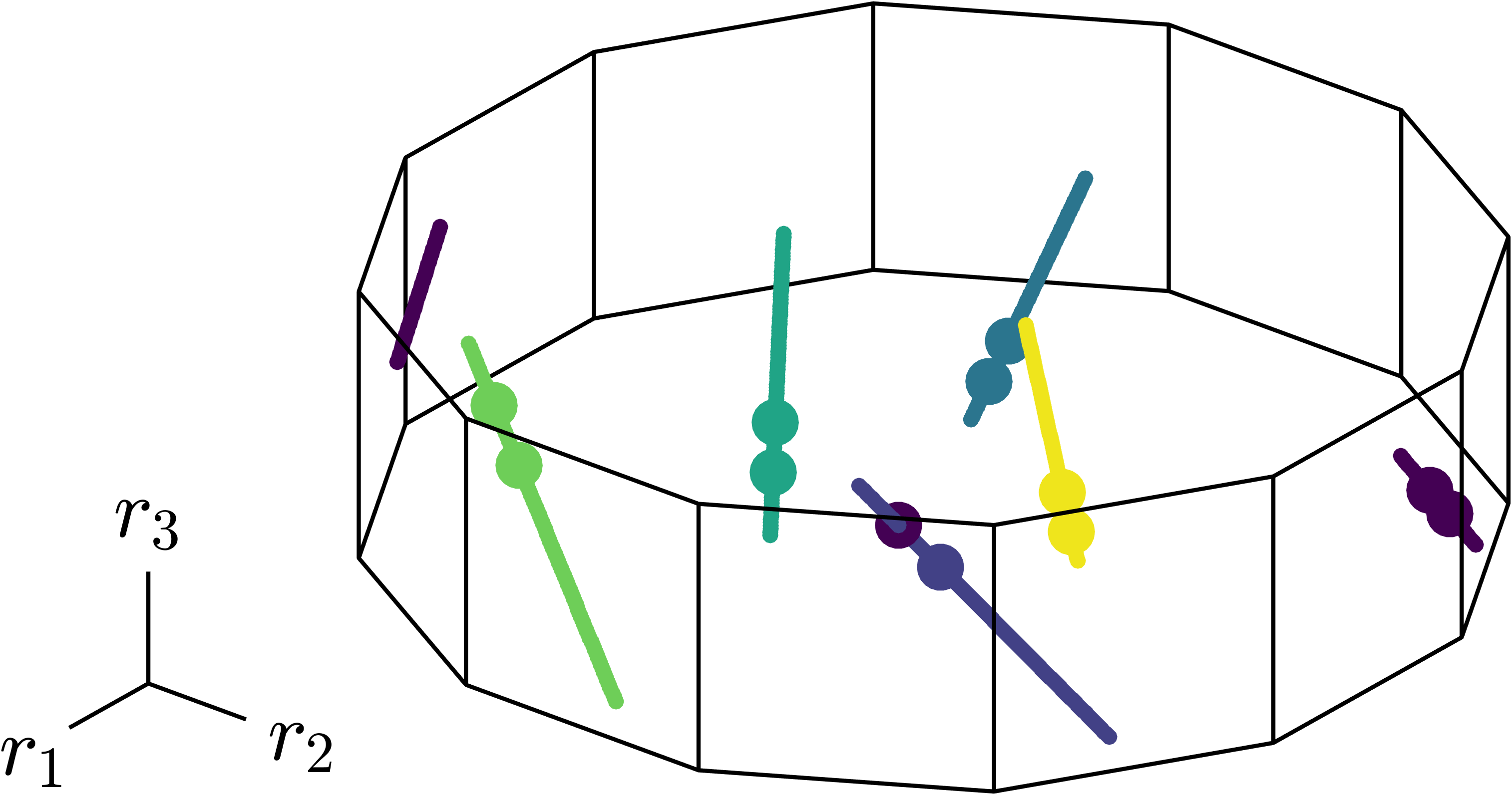}
        \label{subfig:alphaknownfibs_all_true}}
    \subfigure[]{%
	\includegraphics[width=0.35\textwidth]{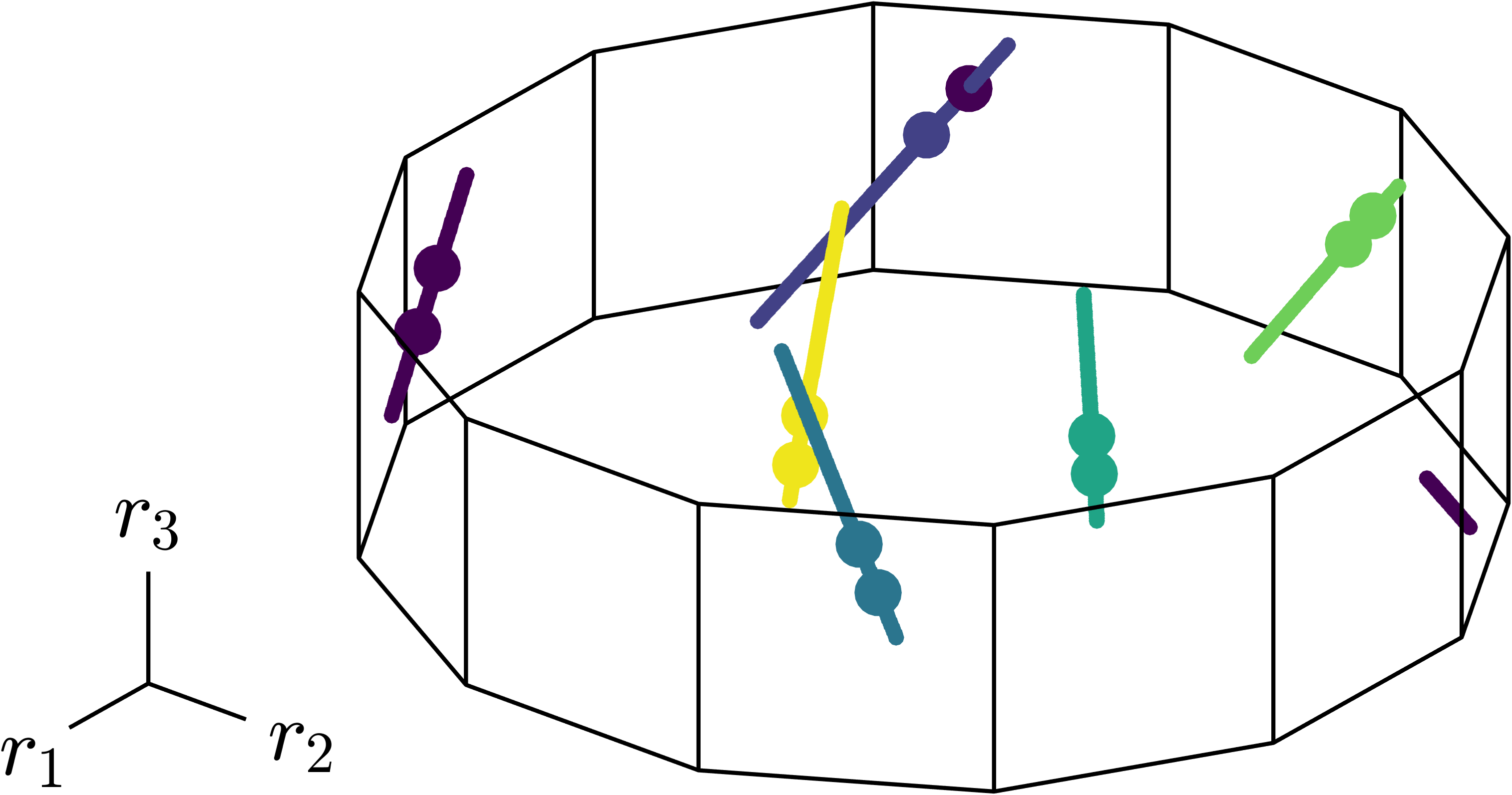}
        \label{subfig:alphaknownfibs_all_pp}}
    \caption{\subref{subfig:alphaknownfibs_all_true} The 12 $\upalpha$ colony variants (dots) that may arise from the random parent $\upbeta$ grain orientation plotted in Figure~\ref{fig:betaknown} plotted in the hexagonal symmetry fundamental region of Rodrigues space, along with their corresponding crystallographic fibers depicting all orientations with the same $c$ axis orientations as their respective deterministic orientations (lines), and \subref{subfig:alphaknownfibs_all_pp} the reflected orientations and fibers. Note that the colors between the two subfigures are corresponding---i.e., a true orientations/fibers are colored the same as their reflected orientations/fibers.}
    \label{fig:alphaknownfibs}
\end{figure}

We present depictions of the variant orientations, their corresponding crystallographic fibers, and the reflected orientations/fibers in Figure~\ref{fig:alphaknownfibs}. These fibers represent the possible fibers which could be measured via PLM for $\upalpha$ colonies that could arise from the parent $\upbeta$ grain presented in Figure~\ref{fig:betaknown}, and would be the starting point of an analysis after measurement. We note that despite the presence of 24 $\upalpha$ colony variants, there exist only 12 unique $\upalpha$ colony fibers (considering both the true variants and their reflections). In other words, two $\upalpha$ colony variants each share the same $c$ axis orientation and thus lie on the same fiber.

\subsection{Deterministic Parent \texorpdfstring{$\upbeta$}{beta} Orientation Based on \texorpdfstring{$\upalpha$}{alpha} Fibers}
\label{subsec:calcbetaori}

\begin{figure}[htbp!]
    \centering
    \subfigure[]{%
	\includegraphics[width=0.3\textwidth]{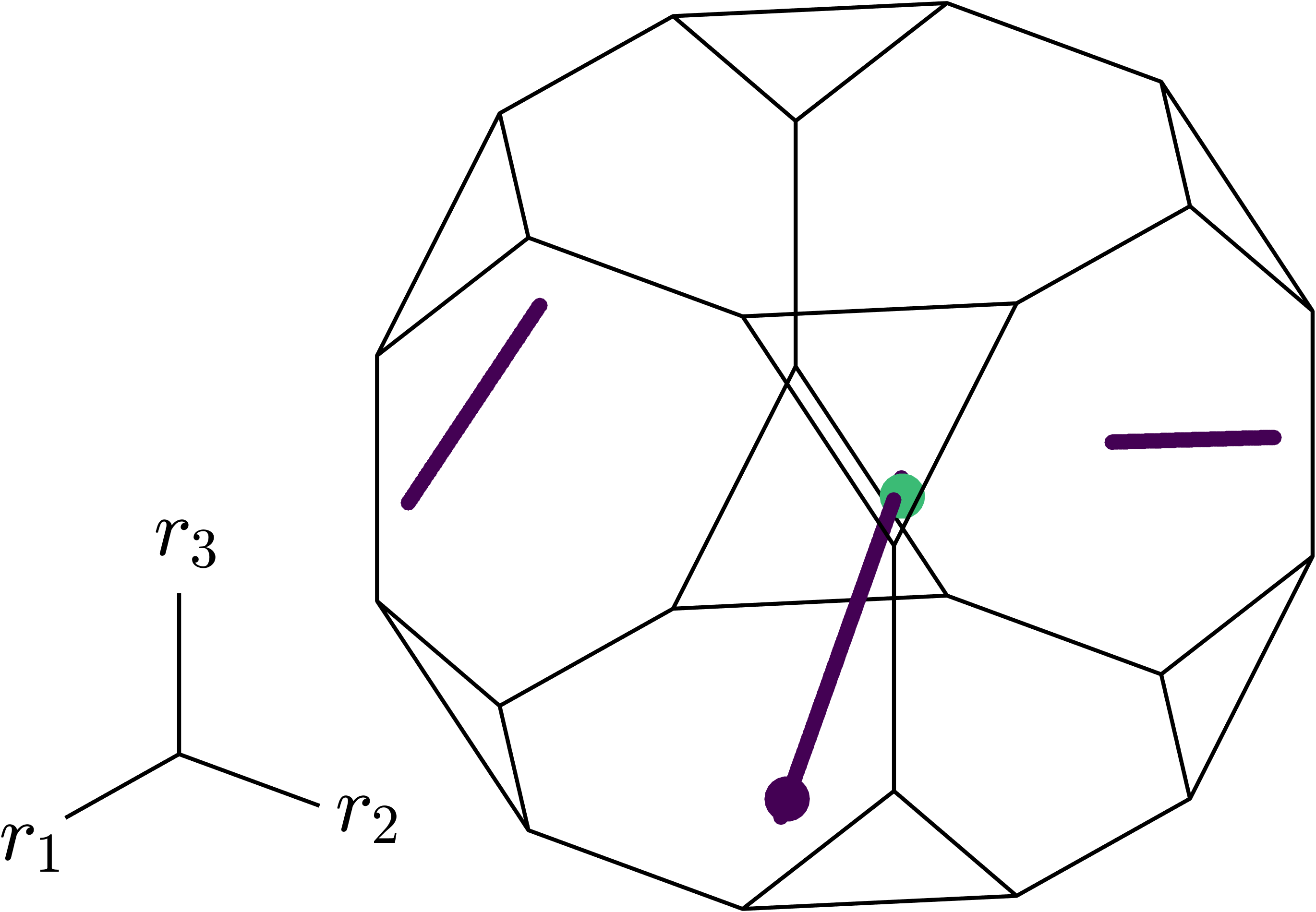}
        \label{subfig:betafibs_1}}
    \subfigure[]{%
	\includegraphics[width=0.3\textwidth]{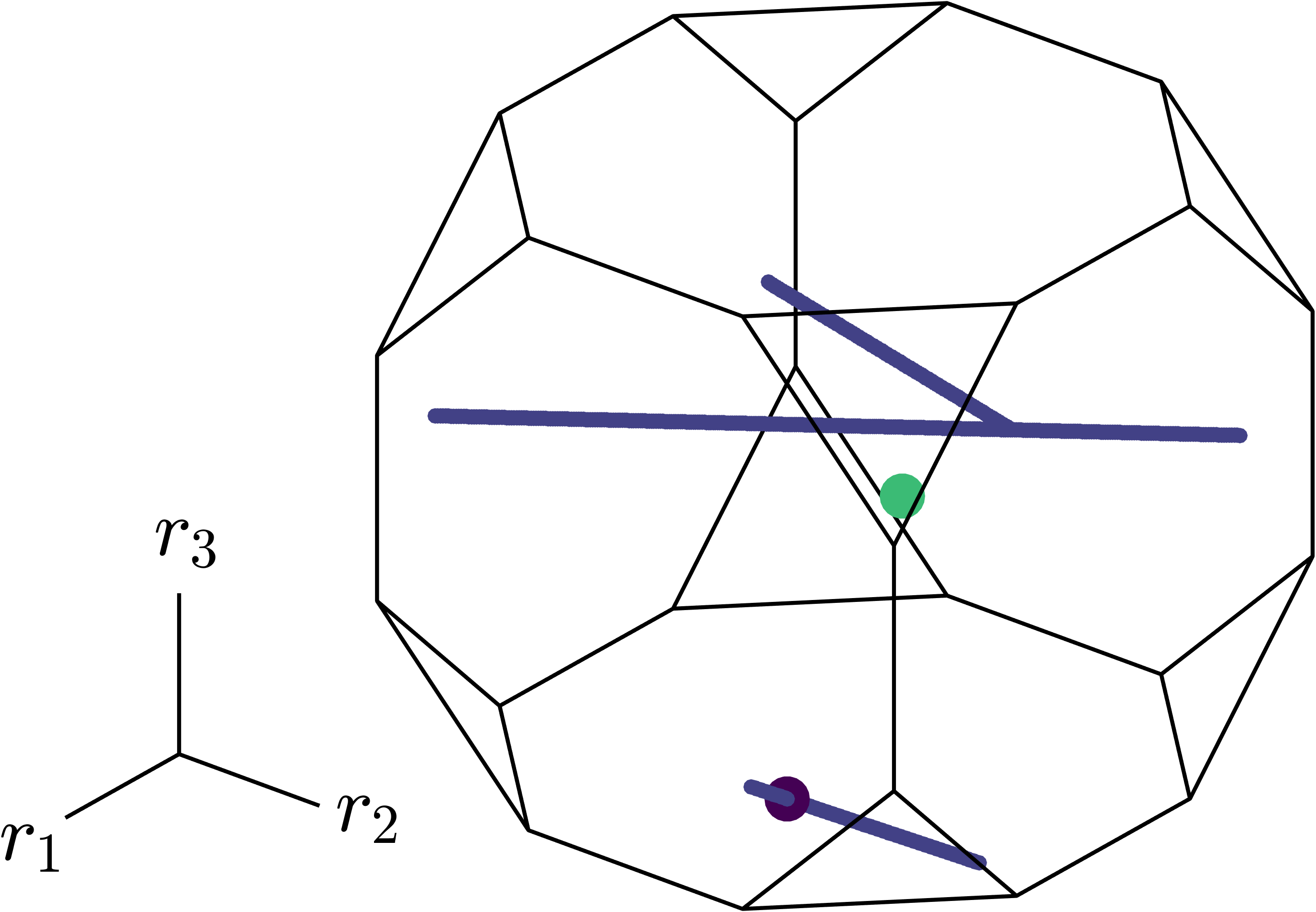}
        \label{subfig:betafibs_2}}
    \subfigure[]{%
	\includegraphics[width=0.3\textwidth]{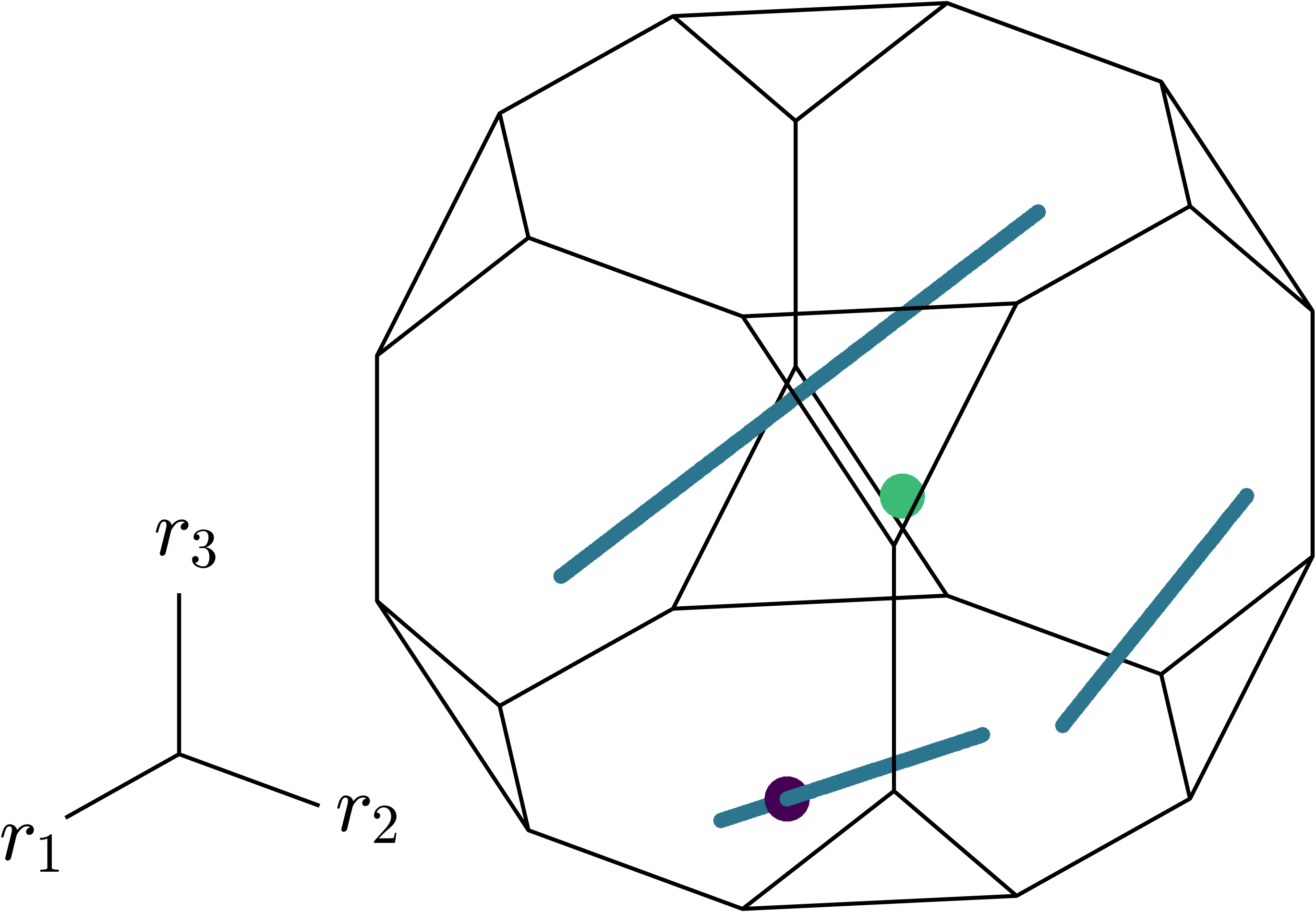}
        \label{subfig:betafibs_3}}
    \subfigure[]{%
	\includegraphics[width=0.3\textwidth]{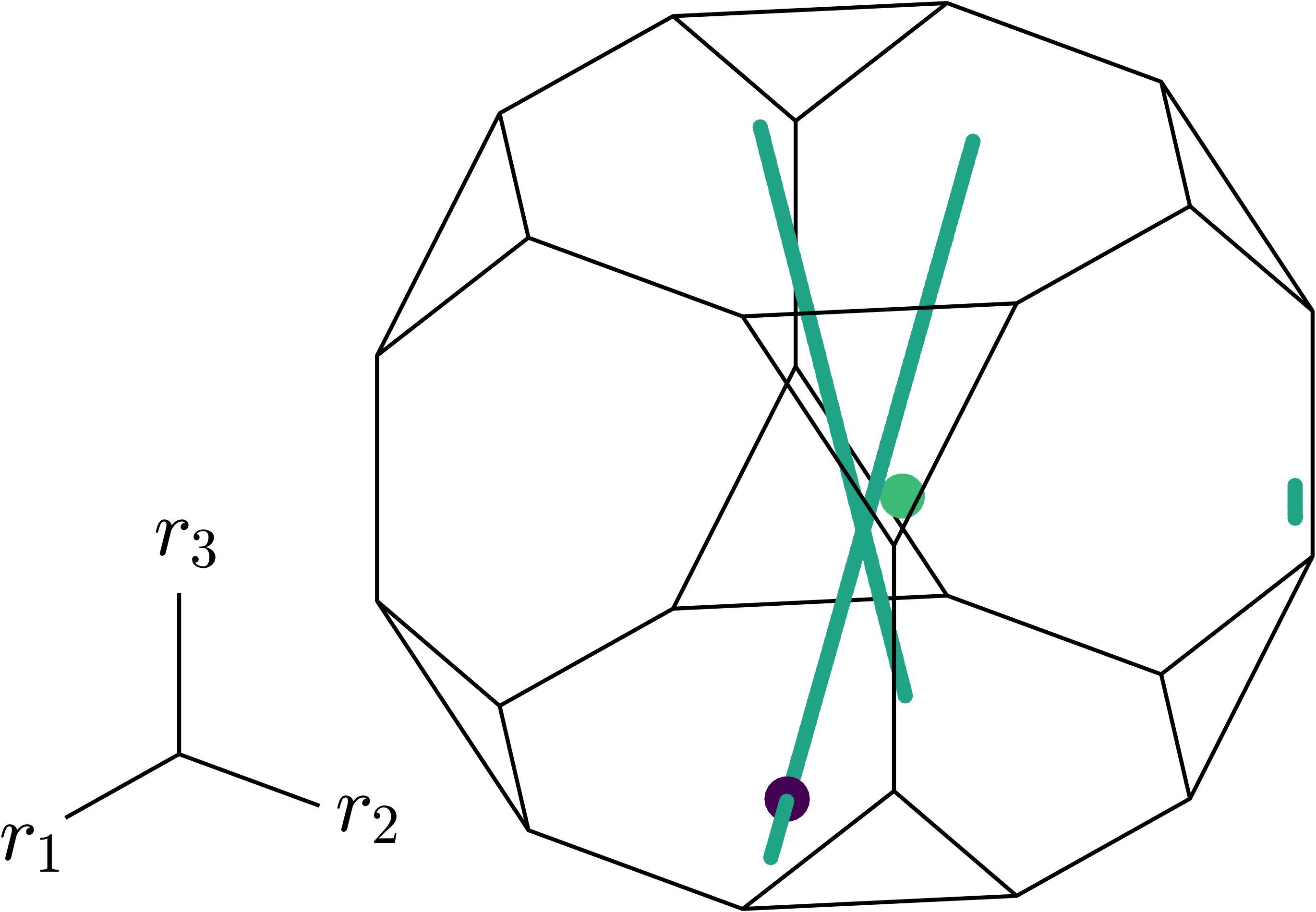}
        \label{subfig:betafibs_4}}
    \subfigure[]{%
	\includegraphics[width=0.3\textwidth]{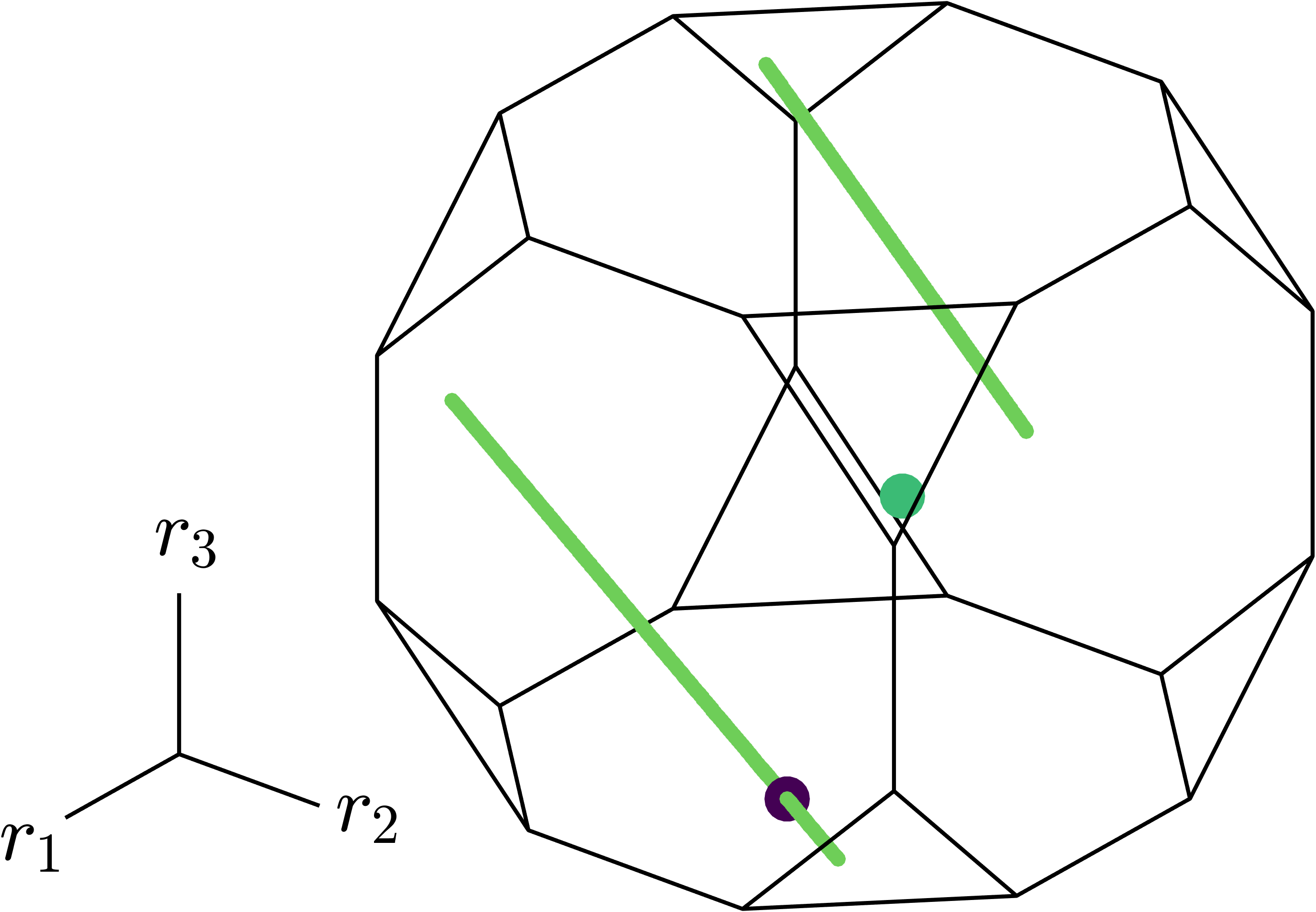}
        \label{subfig:betafibs_5}}
    \subfigure[]{%
	\includegraphics[width=0.3\textwidth]{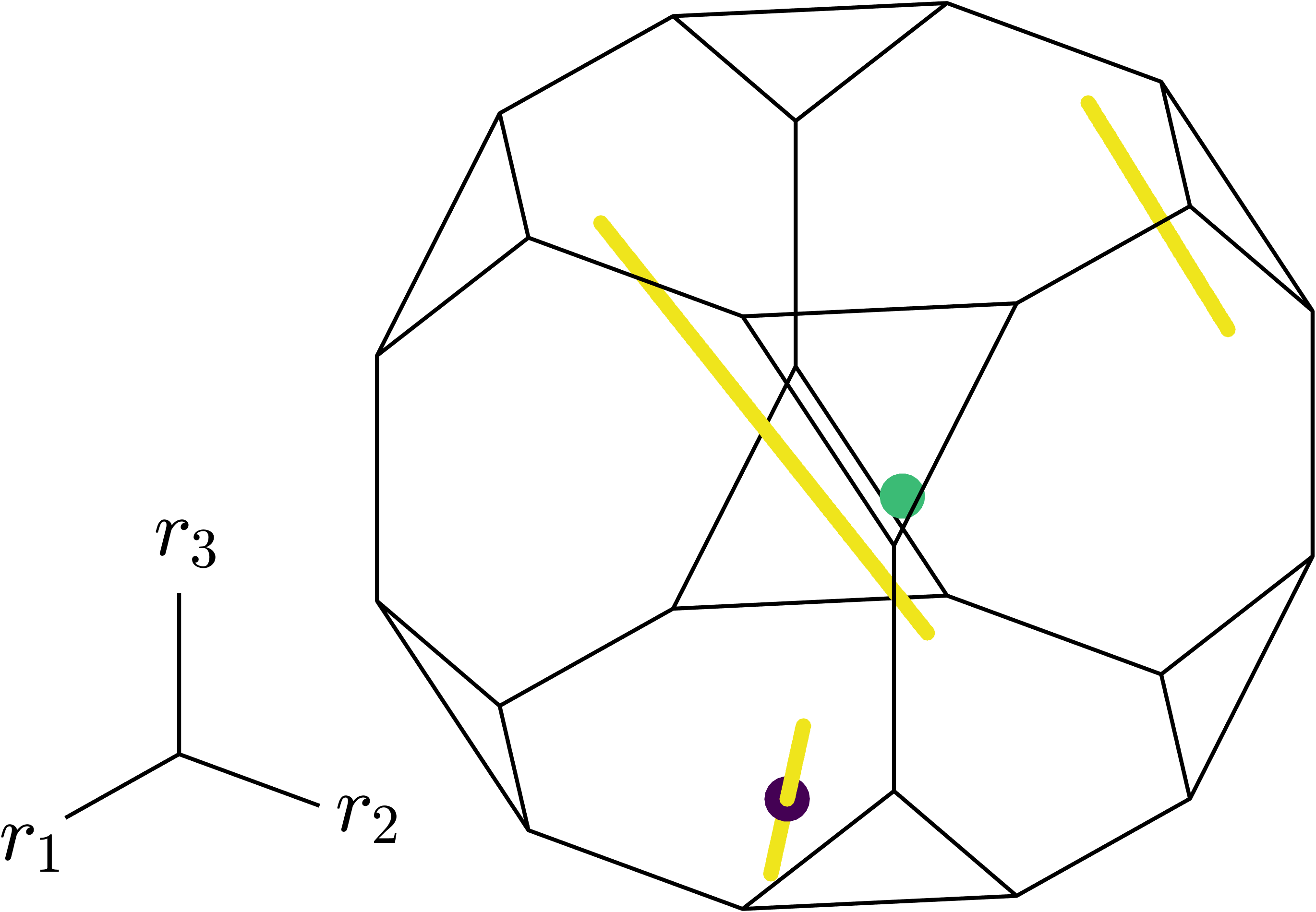}
        \label{subfig:betafibs_7}}
    \caption{Sets of parent $\upbeta$ grain fibers (lines) from which each of the six true $\upalpha$ colony fiber presented in Figure~\ref{fig:alphaknownfibs} could have arisen, plotted in the cubic symmetry fundamental region of Rodrigues space. Additionally, the point of mutual intersect among the each sets of parent $\upbeta$ grain fibers is plotted as a dark dot (compare to the true $\upbeta$ orientation in Figure~\ref{fig:betaknown}), while the reflected orientation is plotted as a light dot. The fibers generally do not intersect the point of the reflected orientation. Note that the colors of the fibers correspond to the true $\upalpha$ colony fibers in Figure~\ref{fig:alphaknownfibs}.}
    \label{fig:betafibs}
\end{figure}

\begin{figure}[htbp!]
    \centering
    \subfigure[]{%
	\includegraphics[width=0.3\textwidth]{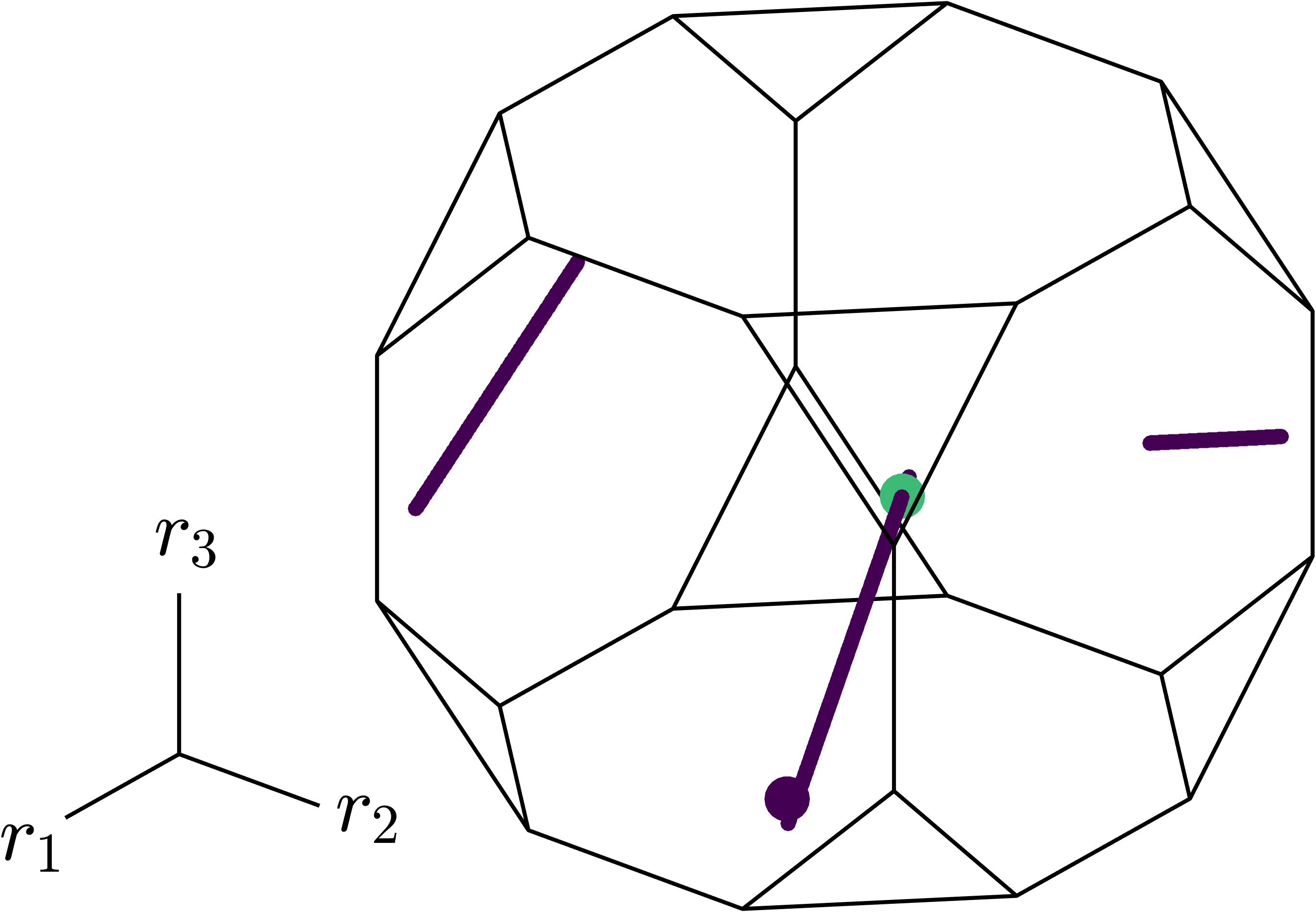}
        \label{subfig:betafibs_13}}
    \subfigure[]{%
	\includegraphics[width=0.3\textwidth]{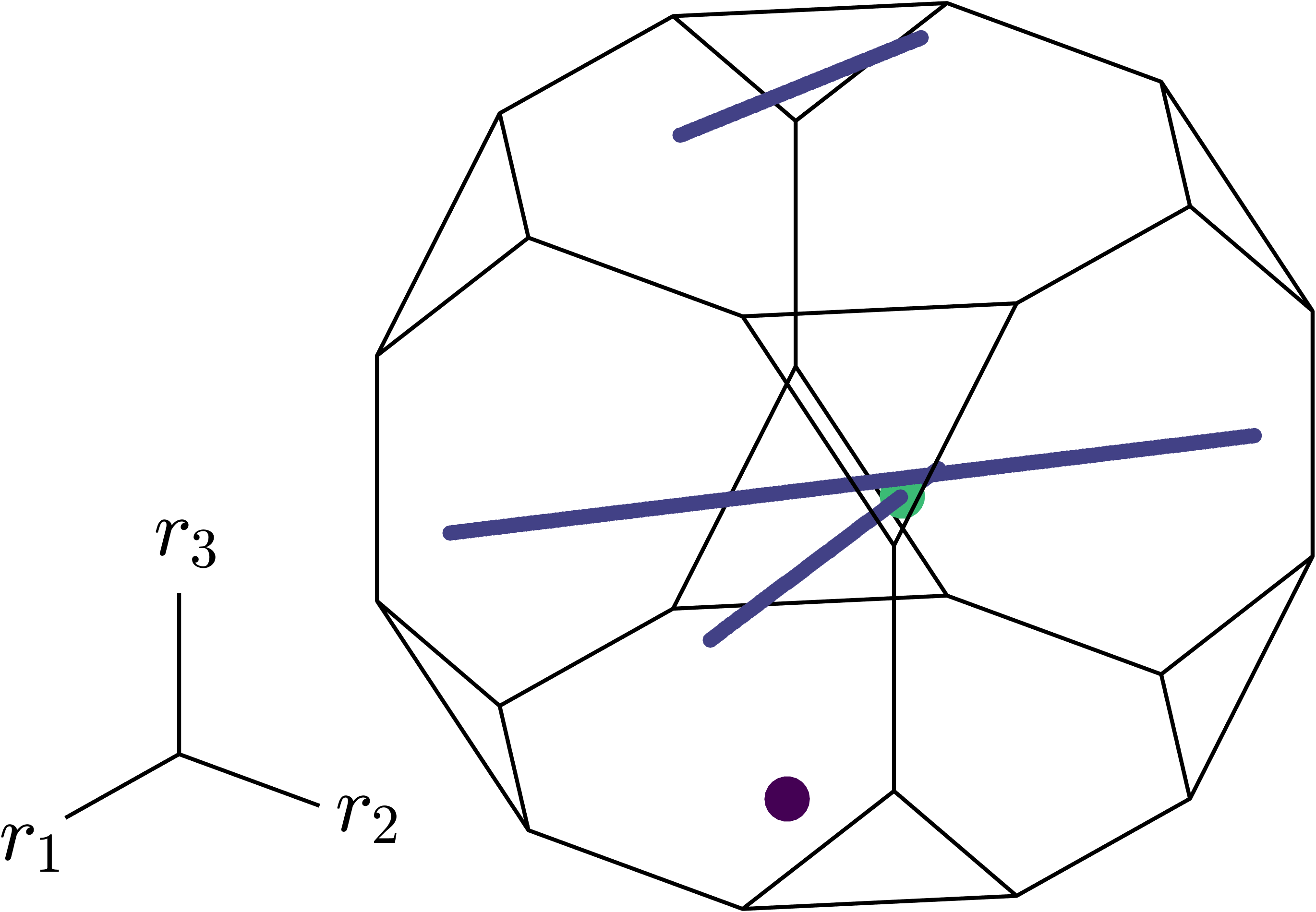}
        \label{subfig:betafibs_14}}
    \subfigure[]{%
	\includegraphics[width=0.3\textwidth]{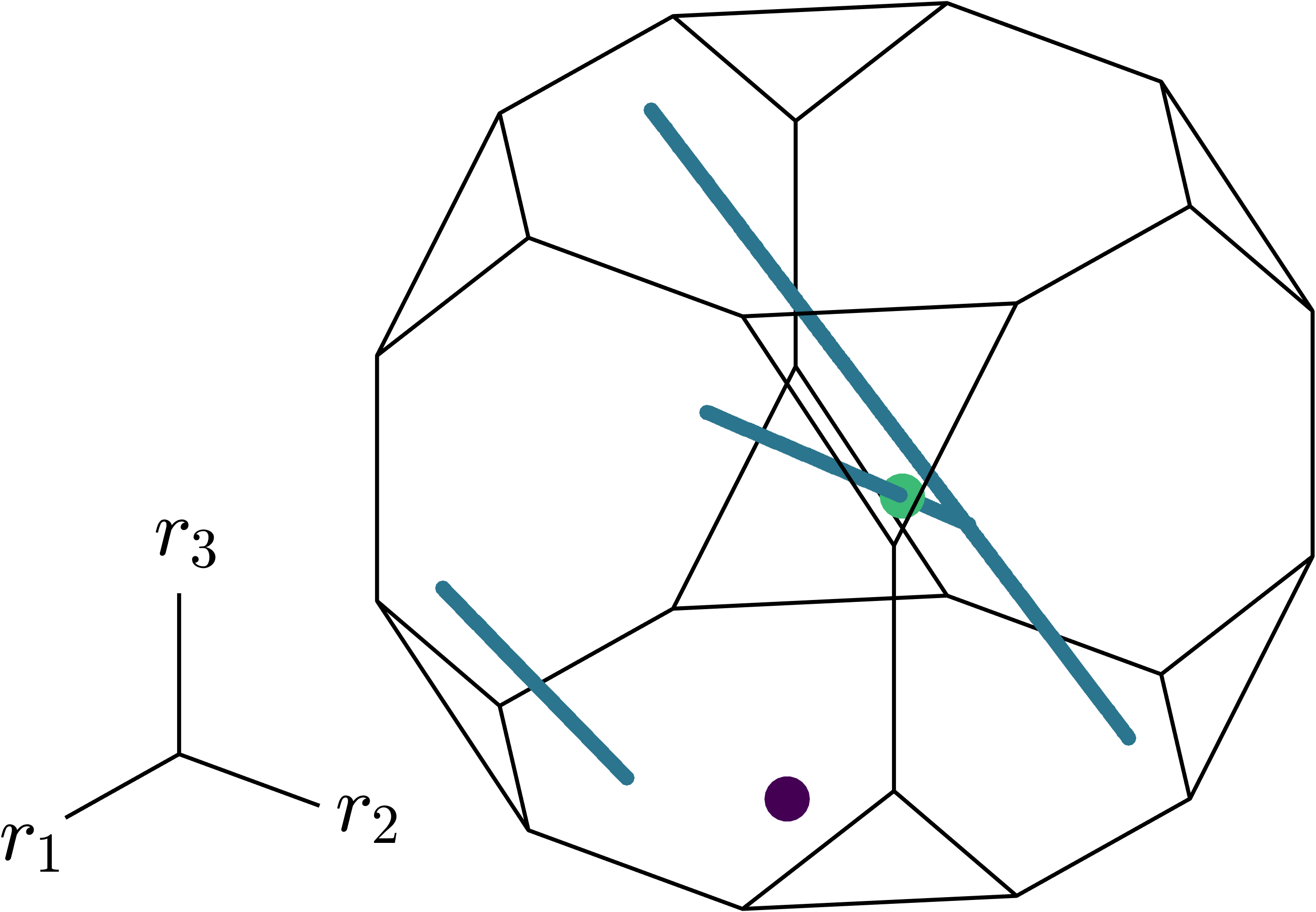}
        \label{subfig:betafibs_15}}
    \subfigure[]{%
	\includegraphics[width=0.3\textwidth]{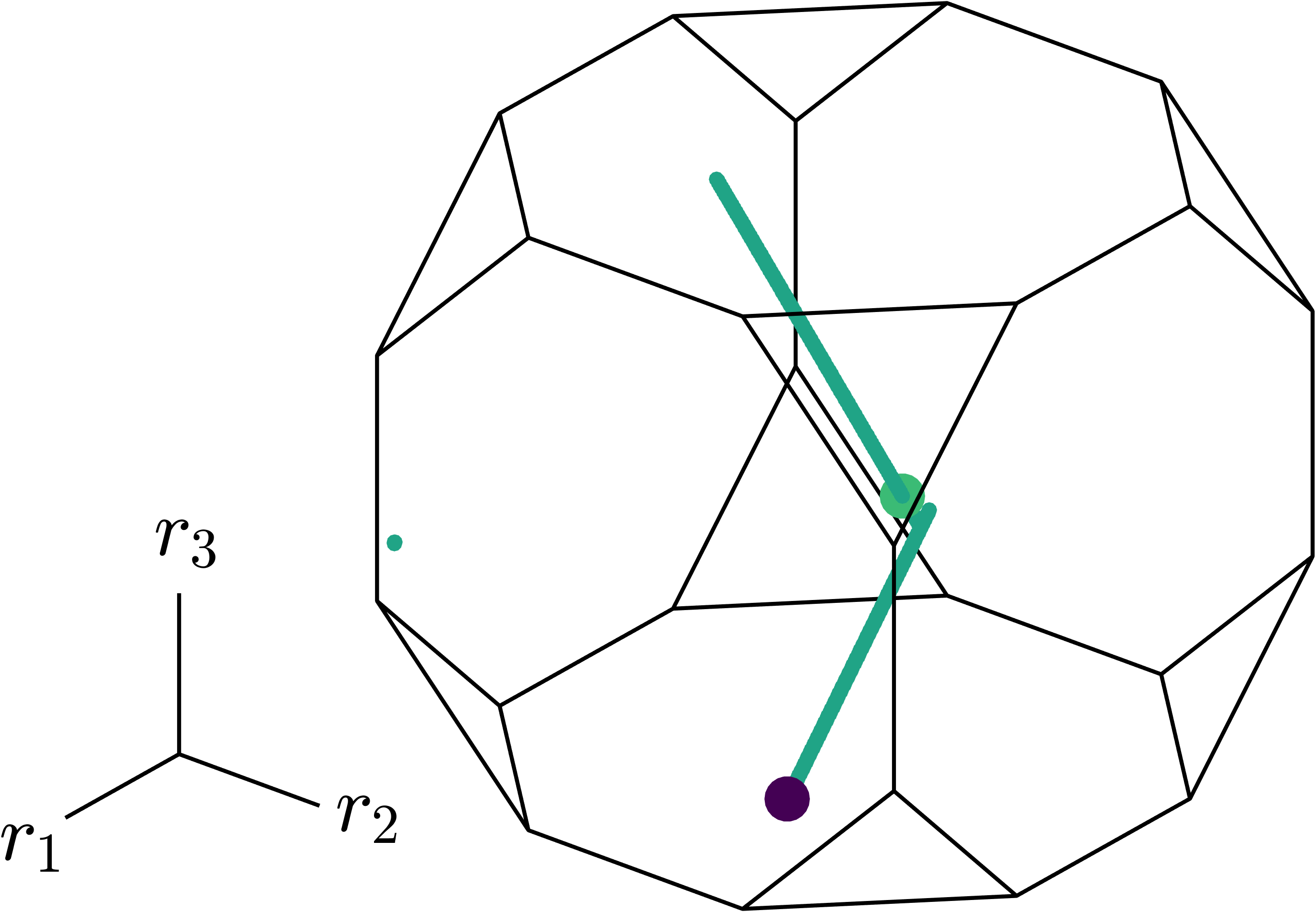}
        \label{subfig:betafibs_16}}
    \subfigure[]{%
	\includegraphics[width=0.3\textwidth]{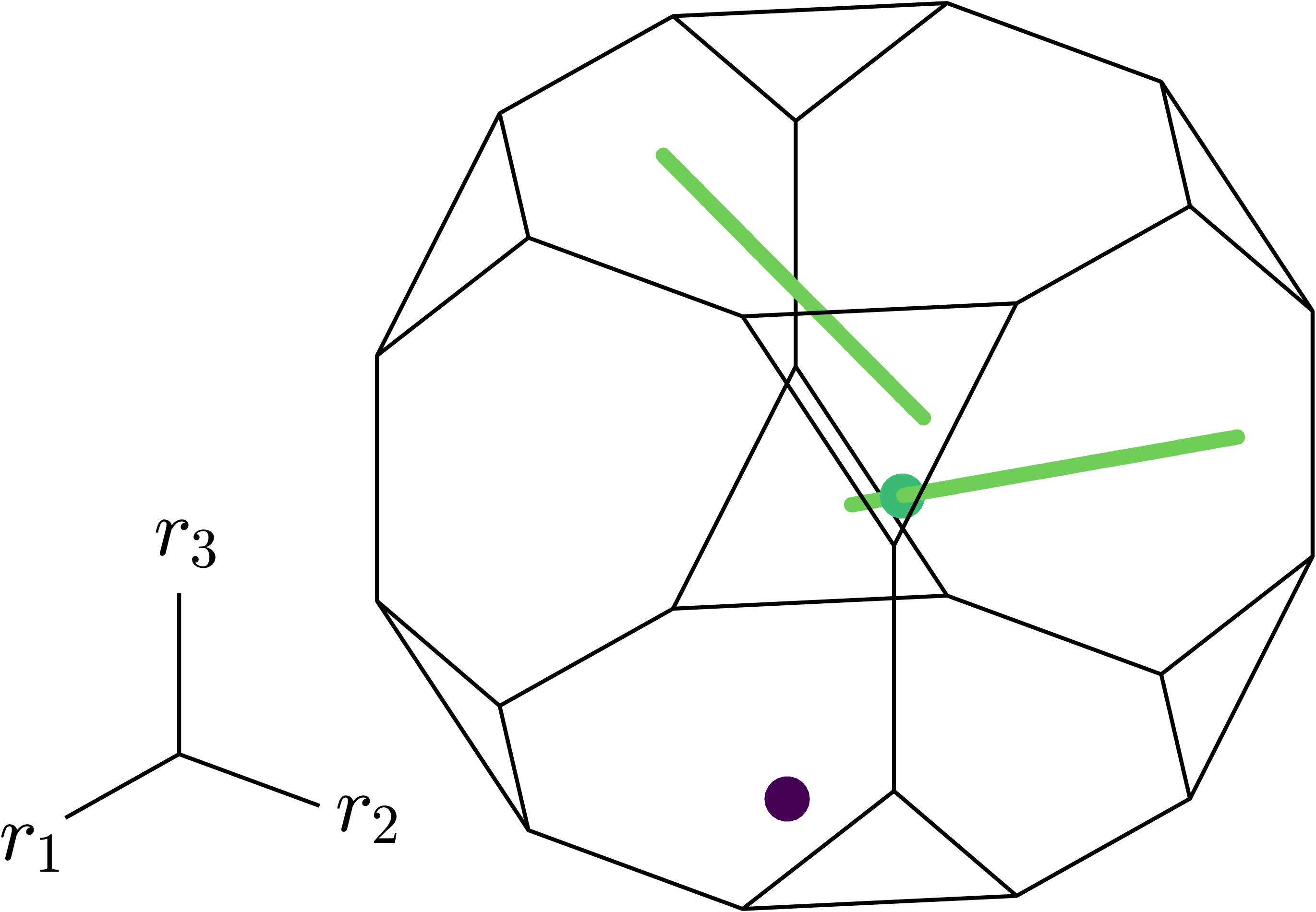}
        \label{subfig:betafibs_17}}
    \subfigure[]{%
	\includegraphics[width=0.3\textwidth]{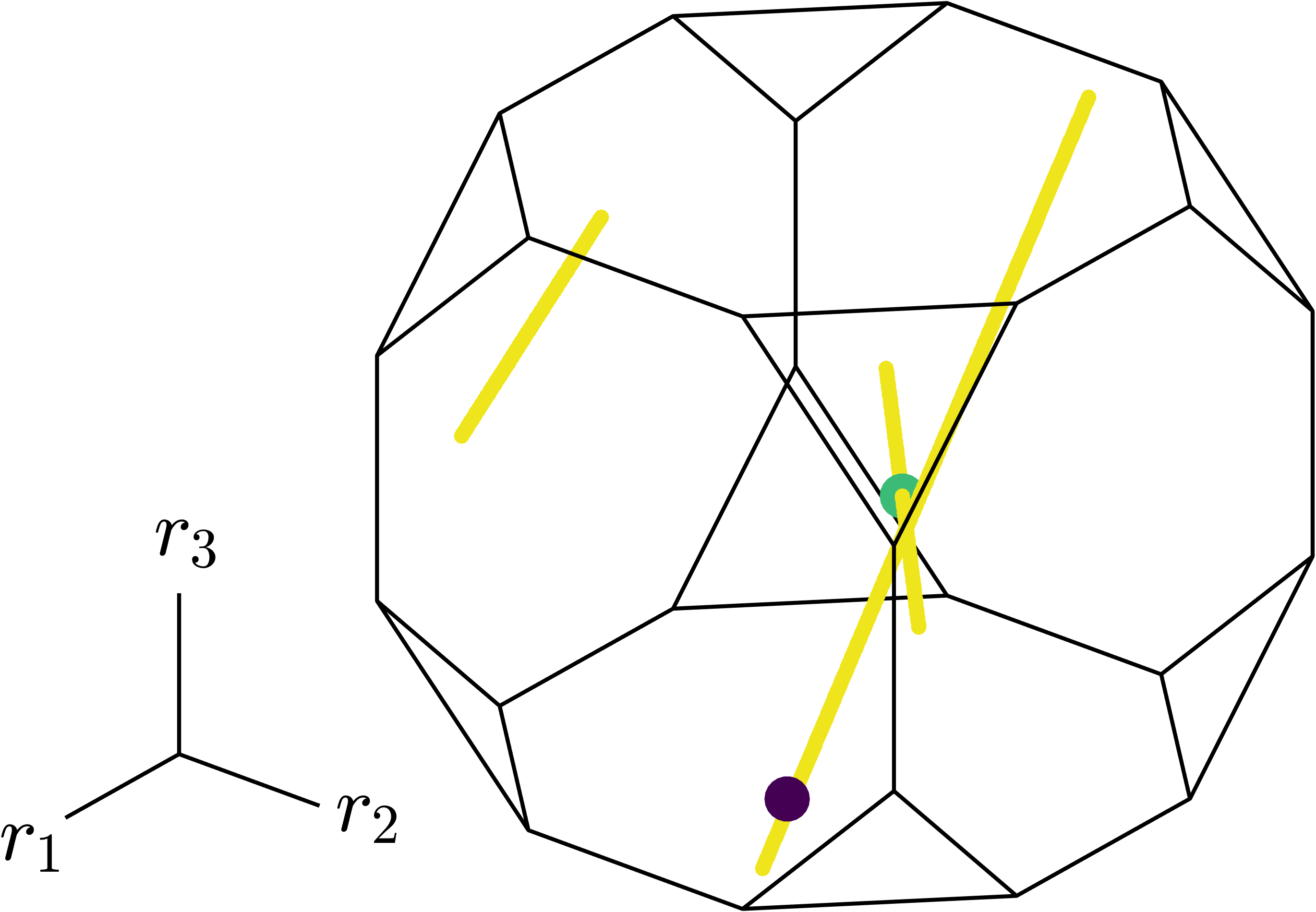}
        \label{subfig:betafibs_19}}
    \caption{Sets of parent $\upbeta$ grain fibers (lines) from which each of the six reflected $\upalpha$ colony fiber presented in Figure~\ref{fig:alphaknownfibs} could have arisen, plotted in the cubic symmetry fundamental region of Rodrigues space. Additionally, the point of mutual intersect among the each sets of parent $\upbeta$ grain fibers is plotted as a light dot, while the true orientation is plotted as a dark dot. The fibers generally do not intersect the point of the true orientation. Note that the colors of the fibers correspond to the reflected $\upalpha$ colony fibers in Figure~\ref{fig:alphaknownfibs}.}
    \label{fig:betafibs2}
\end{figure}

Recall that each $\upalpha$ colony orientation can arise from six possible parent $\upbeta$ grain orientations (Equation~\ref{eq:betatoalpha}). From the set of orientations which lie along the $\upalpha$ colony fibers shown in Figure~\ref{fig:alphaknownfibs}, we calculate the possible parent $\upbeta$ grain variants from which they may have arisen, which leads to a set of parent $\upbeta$ fibers from which each $\upalpha$ colony fiber may have descended. To demonstrate, we calculate the possible parent $\upbeta$ grain fibers for each of the 6 true $\upalpha$ colony fibers, which we plot in Figure~\ref{fig:betafibs}, as well as for the 6 reflected $\upalpha$ colony fibers, which we plot in Figure~\ref{fig:betafibs2}. We note that the only point of common intersect among the six sets of $\upbeta$ fibers deduced from the six true $\upalpha$ variant fibers is the initial parent $\upbeta$ grain orientation (i.e., our demonstrative starting point, presented in Figure~\ref{fig:betaknown}), while the only point of common intersect among the six sets of $\upbeta$ fibers deduced from the six reflected $\upalpha$ variant fibers is a \SI{180}{\degree} rotation of the true parent $\upbeta$ grain orientation. While in this example we know the true orientations and fibers, we contend that---since PLM is unable to discern between the true $\upalpha$ colony fibers and their reflections---the consequence here is that we cannot determine which parent $\upbeta$ grain orientation is true and which is itself a reflected orientation.

\subsection{Quasi-Deterministic \texorpdfstring{$\upalpha$}{alpha} Orientations}
\label{subsec:quasideterministic}

With the possibility that the parent $\upbeta$ grain orientation can be determined to one of two orientations based on the $\upalpha$ colony fibers measured via PLM and their reflections, we can calculate the 24 deterministic orientations of the $\upalpha$ colonies which may transform from the two deterministic parent $\upbeta$ grain orientations. We again refer to Figure~\ref{fig:alphaknownfibs}, which displays these actual $\upalpha$ colony orientations plotted on the fibers. We note that each of the $\upalpha$ colony fibers has two variants which lie along the fiber---i.e., two variants each share the same $c$ axis orientation. Thus, there exists \emph{four} orientations which share the same $c$ axis \emph{or a reflected $c$ axis} (between which PLM cannot distinguish). Thus, we can at best determine the full orientation of the $\upalpha$ colony to one of four orientations (i.e., a quasi-deterministic calculation).

\section{Crystal Plasticity Finite Element Method}
\label{sec:cpfem}

In this study, we utilize the CPFEM software package FEPX~\citep{fepxweb,neperfepx}. FEPX is a free and open-source CPFEM solver which is MPI parallelized, allowing for its deployment on large-scale computational clusters. FEPX is thus capable of high-throughput simulations of the elastic-plastic deformation response of high-fidelity representations of polycrystalline microstructures. FEPX assumes quasi-static, ductile, isothermal deformation, and consequently the models it employs reflects these assumptions. As FEPX uses widely-accepted models employed in a well-established solution approximation scheme (i.e., a non-linear finite element solver), we present here a truncated description of the most important portions of the model for brevity. We refer the reader to~\citep{fepxarxiv} for a full description of the kinematics, models, and finite element implementation. The models below are considered locally at points within each finite element in a polycrystalline mesh.

We employ Hooke's law to relate the Kirchhoff stress tensor, $\boldsymbol{\tau}$, to the elastic strain tensor, ${\boldsymbol\epsilon}^{e}$:
\begin{equation}
    \label{eq:hooke}
    \boldsymbol{\tau} = {\textbf C} \left( {\bf r} \right){\boldsymbol\epsilon}^{e} \quad ,
\end{equation}
where ${\textbf C}$ is the fourth-order anisotropic stiffness tensor. The stiffness tensor reflects the symmetry of the crystal~\citep{bower,nye}, and is orientation dependent, written above as a function of the orientation parameterized as a Rodrigues vector, ${\bf r}$.

For modeling plastic deformation, we consider slip kinetics via a rate-dependent restricted-slip power law~\citep{nemat1986rate} model:
\begin{equation}
    \label{eq:gamma_dot}
    \dot{\gamma}^k= \dot{\gamma}_0 \frac{\tau^k}{\tau_c^k} \left| \frac{\tau^k}{\tau_c^k} \right|^{\frac{1}{m}-1} \quad .
\end{equation}
In this formulation, $\dot{\gamma}$ is the current shear rate on the $k$-th slip system, $\dot{\gamma}_{0}$ is the fixed-state strain rate scaling coefficient, $\tau$ is the resolved shear stress, $\tau_c$ is the critical resolved shear stress, and $m$ is a power parameter controlling rate dependency.

We employ a saturation-style hardening model to evolve the value of the critical resolved shear stress as plasticity evolves:
\begin{equation}
    \label{eq:hardening}
    \dot{\tau}_c^{k} =  h_{0} \left( \frac{\tau_{s}^{k}-\tau_c^{k}}{\tau_{s}^{k}-\tau_{0}^{k}} \right) \dot{\Gamma} \quad .
\end{equation}
Here, $h_0$ is the fixed-state hardening scaling coefficient, $\tau_{s}$ is the saturation value of the critical resolved shear stress, and $\dot{\Gamma}$ is the sum of the absolute value of all shearing rates at the material point (i.e., sum over $k$). We assume that the slip strengths evolve isotropically. For the case of HCP crystals, which have disparate initial strengths between the slip families (basal, prismatic, and pyramidal), the isotropic assumption maintains the initial ratio of the slip strengths through deformation---i.e., the single crystal yield surface retains its shape in stress space as the slip systems harden.

Finally, we model the reorientation of the crystal as the rate change of the orientation, again parameterized as the Rodrigues vector:
\begin{equation}
    \dot{{\textbf r}} = \frac{1}{2} \left( \boldsymbol{\omega} + \left( \boldsymbol{\omega} \cdot {\textbf r} \right){\textbf r} + \boldsymbol{\omega} \times {\textbf r} \right) \quad .
\end{equation}
Here, $\boldsymbol{\omega}$ is the lattice spin vector, described in full in~\citep{fepxarxiv}, itself a function of the amount of local plastic plastic shearing.
\end{appendices}

\section*{Data Availability}
The raw/processed data required to reproduce these findings cannot be shared at this time due to technical or time limitations.

\bibliography{bibliography}
\bibliographystyle{elsarticle-num}






\end{document}